\documentclass[11pt]{article}
\usepackage{amsfonts,amsmath,amssymb}
\usepackage{enumerate}
\usepackage[hidelinks]{hyperref}
\usepackage{bbm}
\usepackage{nicefrac}
\usepackage[all]{xy}
\usepackage{graphicx}
\usepackage{bm}
\usepackage{makecell}
\usepackage[table]{xcolor}

\usepackage{upgreek}
\usepackage{slashed}

\usepackage{float}			
\usepackage{lscape}                 

\usepackage{booktabs}

\newcommand{\be}{\begin{equation}}
\newcommand{\ee}{\end{equation}}

\newcommand{\bea}{\begin{eqnarray}}
\newcommand{\eea}{\end{eqnarray}}

\newcommand{\bes}{\begin{subequations}}
\newcommand{\ees}{\end{subequations}}

\newcommand{\cN}{{\cal N}}

\newcommand{\cA}{{\cal A}}

\def\sst#1{{\scriptscriptstyle #1}}
\def\oneone{\rlap 1\mkern4mu{\rm l}}

\def\0{{\sst{(0)}}}
\def\1{{\sst{(1)}}}
\def\2{{\sst{(2)}}}
\def\3{{\sst{(3)}}}
\def\4{{\sst{(4)}}}
\def\5{{\sst{(5)}}}
\def\6{{\sst{(6)}}}
\def\7{{\sst{(7)}}}
\def\8{{\sst{(8)}}}

\def\cA{{{\cal A}}}

\def\cV{{{\cal V}}}

\newcommand{\vol}{\textrm{vol}}

\allowdisplaybreaks  

\usepackage{multirow}
\usepackage{rotating}

\newcommand{\ba}{\begin{align}}
\newcommand{\ea}{\end{align}}

\newcommand{\bse}{\begin{subequations}}
\newcommand{\ese}{\end{subequations}}

\allowdisplaybreaks

\newcommand{\dd}{\text{d}}

\newcommand{\rep}[1]{\mathbf{#1}}

\newcommand{\GLg}[1]{\mathrm{GL}( #1 )}

\newcommand{\Ex}[1]{\mathrm{E}_{#1(#1)}}

\usepackage{lipsum}

\newlength\Colsep
\newcommand{\uM}{\underline{M}}
\newcommand{\uN}{\underline{N}}
\newcommand{\uP}{\underline{P}}
\newcommand{\uQ}{\underline{Q}}
\newcommand{\uR}{\underline{R}}
\newcommand{\uS}{\underline{S}}

\newcommand{\reals}{\mathbb{R}}

\begin{document}

\makeatletter
\renewcommand{\theequation}{\thesection.\arabic{equation}}
\@addtoreset{equation}{section}
\makeatother

\begin{titlepage}

\begin{flushright}
IFT-UAM/CSIC-26-1
%
\end{flushright}

\vspace{20pt}

   \begin{center}
   \baselineskip=16pt

   \begin{Large}\textbf{
\hspace{-18pt} Maximal trombone supergravity from wrapped M5-branes
}
   \end{Large}

\vspace{25pt}

{\large Martín Pico$^{1,2}$ and  Oscar Varela$^{3,1}$ }

\vspace{30pt}

	\begin{small}

{\it $^1$  Instituto de F\'\i sica Te\'orica UAM/CSIC, 28049 Madrid, Spain} 

	\vspace{12pt}

	{\it $^2$  Departamento de F\'\i sica Te\'orica, Universidad Aut\'onoma de Madrid, \\
	Cantoblanco, 28049 Madrid, Spain} 

	\vspace{12pt}
          
   {\it $^3$ Department of Physics, Utah State University, Logan, UT 84322, USA} 
		
	\end{small}

\vskip 70pt

\end{center}

\begin{center}
\textbf{Abstract}
\end{center}

\begin{quote}

A new family of maximal supergravities in four dimensions, involving gaugings of the trombone scaling symmetry, has been recently introduced. Using exceptional generalised geometry, we show some supergravities in this class to arise by consistent truncation of $D=11$ supergravity. The seven-dimensional reduction manifold is locally equivalent to the topologically-twisted internal manifold of the AdS$_4$ geometries that arise near the horizon of M5-branes wrapped on supersymmetric three-cycles of special holonomy manifolds. The dimensional reduction involves a mixture of conventional and generalised Scherk-Schwarz prescriptions, and provides the first maximally supersymmetric consistent truncation to four dimensions in the context of the M5-brane.

\end{quote}

\vfill

\end{titlepage}

\tableofcontents



\section{Introduction}


In \cite{Pico:2025cmc}, we constructed a family of gaugings of maximal supergravity in four dimensions \cite{Cremmer:1979up,deWit:2007mt,LeDiffon:2008sh,LeDiffon:2011wt} that can be regarded as a generalisation of specific dyonic CSO gaugings of the type discussed in \cite{DallAgata:2011aa,Dall'Agata:2014ita}. The gauge groups, TCSO$(p,q,r; \mathrm{N})$, of these supergravities contain $\textrm{CSO} (p,q,r) \equiv \textrm{SO} (p,q)  \ltimes \mathbb{R}^{(p+q)r}$, with $p+q+r=5$, and a three-dimensional group $G_3$ of Bianchi type $\mathrm{N}$, as subgroups. In a convenient duality frame, $\textrm{SO} (p,q)$ is gauged electrically, $G_3$ magnetically, and $\mathbb{R}^{(p+q)r}$ dyonically. If $G_3$ is unimodular, the $\textrm{TCSO} (p,q,r; \mathrm{N})$ theories simply reduce to concrete gaugings in the class of \cite{DallAgata:2011aa,Dall'Agata:2014ita}. For non-unimodular $G_3$, the $\textrm{TCSO} (p,q,r; \mathrm{N})$ theories are new, and involve a gauging of the $\mathbb{R}^+$ trombone scaling symmetry of ungauged $D=4$ $\cN=8$ supergravity, along with a gauging of a subgroup of the ungauged duality group, E$_{7(7)}$. These TCSO$(p,q,r; \mathrm{N})$ theories are the four-dimensional counterparts of a similar class of $D=5$ $\cN=8$ gauged supergravities recently introduced in \cite{Bhattacharya:2024tjw,Varela:2025xeb}.

It was also argued in \cite{Pico:2025cmc} that a specific theory in this class, TCSO$(5,0,0; \mathrm{V})$, should be related to configurations of M5-branes wrapped on certain supersymmetric three-dimensional cycles, $\Sigma_3$, of special holonomy manifolds, of the type discussed in \cite{Acharya:2000mu,Gauntlett:2006ux} following \cite{Maldacena:2000mw}, see \cite{Gauntlett:2003di} for a review. The submanifold $\Sigma_3$ is either a special lagrangian (SLAG) three-cycle or an associative three-cycle (A3C), depending on whether the special holonomy manifold the M5-branes are put in is a Calabi-Yau three-fold or a seven-dimensional manifold of G$_2$-holonomy \cite{Acharya:2000mu,Gauntlett:2006ux}. Also, $\Sigma_3$ should be endowed with an Einstein metric of negative curvature so that $\Sigma_3$ is identified with the hyperboloid $H^3$, possibly compactified, $H^3/\Gamma$, with a discrete group of isometries $\Gamma$. The construction of \cite{Acharya:2000mu,Gauntlett:2006ux} can be extended to let $\Sigma_3$ still have an Einstein metric, but of positive or vanishing curvature, thereby identifying $\Sigma_3$ with the three-sphere, $S^3$, or the three-torus, $T^3$. A large number $N$ of M5-branes in this configuration develops a supersymmetric anti-de Sitter, $\textrm{AdS}_4 \times (\Sigma_3 \rtimes S^4)$, throat near its horizon that solves, for $\Sigma_3 = H^3$ \cite{Acharya:2000mu,Gauntlett:2006ux}, the field equations of $D=11$ supergravity \cite{Cremmer:1978km}. Here, we have used the symbol $\rtimes$ to denote that the four-sphere $S^4$ is fibred over $\Sigma_3$, as a consequence of supersymmetry being implemented via the topological twist of \cite{Witten:1988ze,Bershadsky:1995qy}.

Three-fold evidence supports the expected relation \cite{Pico:2025cmc} of $D=4$ $\cN=8$ TCSO$(5,0,0;\mathrm{V})$ supergravity to the $\textrm{AdS}_4 \times (\Sigma_3 \rtimes S^4)$ solutions of \cite{Acharya:2000mu,Gauntlett:2006ux}. Firstly, the SO(5) and the type V $G_3$ subgroups of the gauge group should respectively originate from the $S^4$ and the $\Sigma_3 \sim H^3$ submanifolds of the internal eleven-dimensional geometry $\textrm{AdS}_4 \times (\Sigma_3 \rtimes S^4)$ (recall that $H^3$ is locally diffeomorphic to the group $G_3$ of Bianchi type V). Secondly, the TCSO$(5,0,0;\mathrm{V})$ supergravity was shown to admit two supersymmetric AdS vacua with the right symmetries, supersymmetries and linearised mass spectra to potentially uplift to the two eleven-dimensional backgrounds $\textrm{AdS}_4 \times (\Sigma_3 \rtimes S^4)$ of \cite{Acharya:2000mu,Gauntlett:2006ux}. Finally, some submaximal $D=4$ supergravity models \cite{Gauntlett:2002rv,Donos:2010ax} were previously known to arise from fully non-linear consistent truncation of $D=11$ supergravity on the seven-dimensional manifolds $\Sigma_3 \rtimes S^4$. These submaximal four-dimensional supergravities \cite{Gauntlett:2002rv,Donos:2010ax} can be retrieved \cite{Pico:2025cmc} as  consistent subsectors of $D=4$ $\cN=8$ TCSO$(5,0,0;\mathrm{V})$-gauged supergravity.

This logic provides a strong smoking gun that the entire $D=4$ $\cN=8$ (tromboneful) TCSO$(5,0,0;\mathrm{V})$ supergravity locally arises from fully non-linear consistent truncation of $D=11$ supergravity \cite{Cremmer:1978km} on the seven-dimensional  M5-brane geometries $\Sigma_3 \rtimes S^4$ of \cite{Acharya:2000mu,Gauntlett:2006ux}, with $\Sigma_3 = H^3$. This is indeed the case. In fact, we first found our $D=4$ supergravity by consistent truncation and, in this paper, we will provide a full account of this construction. Similarly, we will recover the (tromboneless) TCSO$(5,0,0;\mathrm{IX})$ and TCSO$(5,0,0;\mathrm{I})$ theories as consistent truncations of $D=11$ supergravity on the seven-dimensional manifolds $\Sigma_3 \rtimes S^4$ with, respectively, $\Sigma_3 = S^3$ and $\Sigma_3 = T^3$. More generally, we will show that the subfamily of $D=4$ $\cN=8$ supergravities with gauge group TCSO$(5,0,0;\mathrm{N})$, with subgroup $G_3$ of arbitrary Bianchi type N, arises from $D=11$ consistent truncation on $G_3 \rtimes S^4$ where the fibration of $S^4$ on $G_3$ is as in the $\Sigma_3 \rtimes S^4$ cases of \cite{Acharya:2000mu,Gauntlett:2006ux}. This truncation goes through consistently even if $G_3$ does not generically admit an Einstein metric. In any case, our construction provides the first $D=11$ truncation to $D=4$ $\cN=8$ gauged supergravity in the context of the M5-brane.

In the previous discussion, as well as in the main text, we have been rather cavalier in leisurely exchanging the compact Riemann surface $\Sigma_3$ that the $D=11$ M5-branes wrap, with the group manifold $G_3$. This is because the local construction in either case is identical. Global aspects, however, are important, particularly in the physically relevant case of compact hyperbolic wrapped cycle, $\Sigma_3 = H^3/\Gamma$, the case with (previously known \cite{Acharya:2000mu,Gauntlett:2006ux}) AdS$_4$ solutions. In this case, the $\Sigma_3 = H^3/\Gamma$ and the Bianchi type V group $G_3$ are still locally identical, but they differ globally: the former is compact while the latter, being non-unimodular, is not. This means that our $D=4$ $\cN=8$ truncations on $\Sigma_3 \rtimes S^4$ only hold locally, and only extend to bonafide global constructions within the supergravity subsectors that do not induce deformations of $\Sigma_3$. As we will see, these are the specific subsectors of $D=4$ $\cN=8$ TCSO$(5,0,0;\mathrm{V})$ supergravity discussed in \cite{Pico:2025cmc}, which were previously obtained \cite{Gauntlett:2002rv,Donos:2010ax} from straight reduction of $D=11$ without passing through the intermediate $D=4$ $\cN=8$ supergravity.

We will build our new consistent truncations using exceptional generalised geometry (ExGG) \cite{Coimbra:2011nw,Coimbra:2011ky,Coimbra:2012af}. Like similar duality-covariant reformulations \cite{Berman:2010is,Hohm:2013pua,Hohm:2013uia}, see \cite{Berman:2020tqn} for a review, of the higher-dimensional supergravities, ExGG makes manifest the $\mathbb{R}^+ \times \mathrm{E}_{7(7)}$ global symmetry of (ungauged) $D=4$ $\cN=8$ supergravity already in eleven-dimensions. While the traditional construction of consistent truncations is a famously intricate and obscure problem, this is greatly streamlined in ExGG \cite{Berman:2012uy,Lee:2014mla,Inverso:2017lrz} by suitably adapting techniques from the conventional approach to supergravity truncations, including Scherk-Schwarz \cite{Scherk:1979zr} and $G$-structure methods \cite{Gauntlett:2007ma}. In the ExGG language, consistent truncations are associated to the existence of generalised $G$-structures \cite{Lee:2014mla,Cassani:2019vcl}, see also \cite{Cassani:2020cod,Josse:2025uro}. In particular, a consistent truncation from $D=11$ to $D=4$ $\cN=8$ supergravity requires the internal seven-dimensional manifold to be equipped with a generalised identity structure (GIS), whose constant instrinsic torsion is identified with the embedding tensor of the maximal four-dimensional supergravity \cite{Lee:2014mla,Cassani:2019vcl}.

We will begin by reviewing the family of $D=4$ $\cN=8$ TCSO$(p,q,r;\mathrm{N})$ supergravities \cite{Pico:2025cmc} in section \ref{sec:4DSugra}, in order to set up conventions. Section \ref{sec:CTFrom11D} contains our central result: the construction in ExGG of a GIS on $G_3 \rtimes S^4$ with constant intrinsic torsion given, as expected, by the embedding tensor of the four-dimensional TCSO$(5,0,0;\mathrm{N})$ theory. In order to do this, we first build a GIS on the direct product manifold $G_3 \times S^4$, and then adapt a twisting prescription given in \cite{Cassani:2019vcl} to obtain the GIS on $G_3 \rtimes S^4$ for the two fibrations associated to the two different, SLAG and A3C, $\textrm{AdS}_4 \times (\Sigma_3 \rtimes S^4)$ M5-brane solutions of \cite{Acharya:2000mu,Gauntlett:2006ux}. These results not only establish the consistency of the truncation of $D=11$ supergravity either on $G_3 \times S^4$ or on the physically more relevant fibred manifolds $G_3 \rtimes S^4$. They also imply, interestingly, that the resulting truncated $D=4$ $\cN=8$ theory is the same, TCSO$(5,0,0;\mathrm{N})$, in all cases. As we will also explain in section \ref{sec:CTFrom11D}, this happens because this type of ExGG twists \cite{Cassani:2019vcl} turn out to be generated, at least in our context and certainly not in an obvious way, by scalars of the $D=4$ $\cN=8$ supergravity. In section \ref{sec:Subsectors} we particularise our general $\cN=8$ truncation to the specific supergravity subsectors considered in \cite{Pico:2025cmc}, in order to make contact with the submaximal truncations of \cite{Gauntlett:2002rv,Donos:2010ax}. Finally, section \ref{sec:Discussion} concludes with further discussion.

\newpage 


\section{$D=4$ $\cN=8$ $\mathrm{TCSO}(p,q,r;\mathrm{N})$-gauged supergravity} \label{sec:4DSugra}


For later reference, let us begin by reviewing some aspects of the family of gaugings of maximal four-dimensional supergravity that we introduced in \cite{Pico:2025cmc}.


\subsection{Field content and embedding tensor} \label{sec:FieldContentET}


We work in the embedding tensor formalism \cite{deWit:2007mt,LeDiffon:2008sh,LeDiffon:2011wt}, see \cite{Trigiante:2016mnt} for a review, where bosons and fermions formally lie in representations of $\mathbb{R}^+ \times \mathrm{E}_{7(7)}$ and $\mathbb{R}^+ \times \mathrm{SU}(8)$, respectively. Here, $\mathbb{R}^+$ is the trombone scaling symmetry, $\mathrm{E}_{7(7)}$ is the duality symmetry group of the ungauged theory \cite{Cremmer:1979up}, and $\mathrm{SU}(8)$ its maximal compact subgroup. We will focus on the bosonic sector. The metric, $g_{\mu \nu}$, gauge fields, $A_\mu{}^{\uM}$, and two-form potentials, $B_{\mu\nu \, \alpha}$, respectively sit in the $\bm{1}_2$, $\bm{56}_1$ and $\bm{133}_2$ of $\mathbb{R}^+ \times \mathrm{E}_{7(7)}$, while the scalars parametrise a coset representative, $\cV_{\uM}{}^{\uN}$, of $\mathrm{E}_{7(7)} / \mathrm{SU}(8)$, with metric $M_{\uM \uN} = (\cV \cV^{\mathrm{T}})_{\uM \uN}$. We use indices $\mu=0,1, 2 , 3$, $\uM= 1 , \ldots , 56$ and $\alpha = 1, \ldots , 133$ to label spacetime vectors, and the fundamental and adjoint representations of $\mathrm{E}_{7(7)}$. The latter indices are respectively raised and lowered with the symplectic form, $\Omega_{\uM\uN}$, of Sp$(56)$  and the Killing-Cartan form, $\kappa_{\alpha \beta}$, of $\mathrm{E}_{7(7)}$. While manifestly $\mathbb{R}^+ \times \mathrm{E}_{7(7)}$ covariant, the above field content is redundant at face value. Duality relations must be imposed that ensure the propagation of only 128 bosonic (and 128 fermionic) degrees of freedom \cite{deWit:2007mt,LeDiffon:2008sh,LeDiffon:2011wt}.

Only in ungauged supergravity are $\mathbb{R}^+ \times \mathrm{E}_{7(7)}$ and $\mathbb{R}^+ \times \mathrm{SU}(8)$ symmetries. In the gauged theory, the (local) symmetry is typically reduced to a subgroup of the former. In the case at hand, we are interested in the family of gauge groups \cite{Pico:2025cmc}
\begin{equation} \label{eq:ProdTCSO}
\textrm{TCSO} (p,q,r;\mathrm{N}) \equiv \big( G_3 \times \textrm{CSO} (p,q,r) \big)   \ltimes \mathbb{R}^{15}  \,  \subset \, \mathbb{R}^+ \times \mathrm{E}_{7(7)} \; , \qquad p+q+r = 5 \; ,
\end{equation}
of dimension 
\begin{equation} \label{eq:dimTCSO}
\textrm{dim} \, \textrm{TCSO} (p,q,r;\mathrm{N}) = \tfrac12 \, (p+q)(p+q-1) + (p+q) \, r + 18 \; ,
\end{equation}
less than 28. In (\ref{eq:ProdTCSO}), $\textrm{CSO} (p,q,r) \equiv \textrm{SO} (p,q)  \ltimes \mathbb{R}^{(p+q)r}$, and $G_3$ is a three-dimensional group of Bianchi type $\mathrm{N}$. If $G_3$ is unimodular, the group (\ref{eq:ProdTCSO}) is strictly contained in $\mathrm{E}_{7(7)}$, and reduces to specific dyonic CSO gaugings of the type discussed in \cite{DallAgata:2011aa,Dall'Agata:2014ita}. For non-unimodular $G_3$, the group $\textrm{TCSO} (p,q,r;\mathrm{N})$ is contained in both factors $\mathbb{R}^+ \times \mathrm{E}_{7(7)}$ and lies outside the class of \cite{DallAgata:2011aa,Dall'Agata:2014ita}. 

A formulation of the gauged supergravity exists with only $\textrm{dim} \, \textrm{TCSO} (p,q,r;\mathrm{N})$ gauge fields active, sitting in the adjoint representation of $\textrm{TCSO} (p,q,r;\mathrm{N})$. However, it is helpful to proceed differently, and formally maintain E$_{7(7)}$ covariance while introducing charges under $\textrm{TCSO} (p,q,r;\mathrm{N})$ by using the so-called embedding tensor  \cite{deWit:2007mt,LeDiffon:2008sh,LeDiffon:2011wt}. We will denote this tensor either by $X_{\uM}$ or by $X_{\uM \uN}{}^{\uP}$, either omitting or displaying its E$_{7(7)}$ representation indices, depending on convenience. The embedding tensor is a constant object in the $\bm{56}_{-1} + \bm{912}_{-1}$ of $\mathbb{R}^+ \times \mathrm{E}_{7(7)}$, which takes values in the Lie algebra of $\mathbb{R}^+ \times \mathrm{E}_{7(7)}$, and governs how the  Lie algebra of the gauge group is embedded in the former. Only equivalence classes of embedding tensors are physically distinct. Namely, the same gauged supergravity can be described by either embedding tensor, $X_{\uM \uN}{}^{\uP}$ or $\tilde{X}_{\uM \uN}{}^{\uP}$, related by an $\mathrm{E}_{7(7)}$ transformation $U_{\uM}{}^{\uN}$ via
\begin{equation} \label{eq:TransXSymbol}
\tilde{X}_{\uM \uN}{}^{\uP} = U_{\uM}{}^{\uQ} \, U_{\uN}{}^{\uR} \, X_{\uQ \uR}{}^{\uS}  \, (U^{-1})_{\uS}{}^{\uP} \; .
\end{equation}
In usual parlance, a choice of embedding tensor within the same duality class (\ref{eq:TransXSymbol}) defines a `duality frame'. Regardless of the duality frame employed to write a $D=4$ $\cN=8$ gauged supergravity, the latter is completely specified by its embedding tensor.

In a convenient duality frame, which will be referred to as the `direct product (DP) frame', the
$D=4$ $\cN=8$ $\textrm{TCSO} (p,q,r;\mathrm{N})$ embedding tensor can be written as 
\begin{equation} \label{eq:ETtromboneSplit}
X_{\uM} = (X_{AB} , X^{AB})= (X_{ab} , X_{ai} , X_{ij} , X^{ab} , X^{ai} , X^{ij}) \; .
\end{equation}
Indices $A=1 , \ldots , 8$ in the intermediate step label the fundamental representation of SL$(8,\mathbb{R})$, while $a=1,2,3$ and $i=4, \ldots , 8$ respectively label the adjoint representation of the $G_3$ and the fundamental of the CSO$(p,q,r)$, $p+q+r=5$, subgroups in (\ref{eq:ProdTCSO}). The different contributions to the embedding tensor (\ref{eq:ETtromboneSplit}) are \cite{Pico:2025cmc}
{\setlength\arraycolsep{0pt}
\begin{eqnarray} \label{eq:ETtrombone}
&& X_{ij} = -2 g_1 \, \eta_{k [i} \, t_{j]}{}^k \; , \qquad 
X_{ai} = g_1 \, t_a{}^j \eta_{ij} -\tfrac{1}{18} g_2 \, f_{ab}{}^{b} \epsilon^{cde} t_{icde}  \; , \qquad 
X_{ab} = X^{ij} = 0  , \quad \\[4pt]
&& 
X^{ai} =  -\frac{1}{2} g_2 \,  \epsilon^{bcd} f_{bc}{}^a \, t_d{}^i    \; , \quad
X^{ab} =  \tfrac12 g_2 \epsilon^{abc} f_{cd}{}^{d} (t_0 - \tfrac43 t_e{}^e)  + g_2 \epsilon^{abc} f_{cd}{}^e ( t_e{}^d - \tfrac13 \delta^d_e t_f{}^f) \; . \nonumber 
\end{eqnarray}
}Here, $t_0$ and $t_\alpha = (t_A{}^B ,  t_{ABCD} ) =  (t_a{}^b , t_a{}^j , t_i{}^b , t_i{}^j , t_{abc\ell}, t_{abk\ell}, t_{ajk\ell} , t_{ijk\ell} )$ respectively are the generators of $\mathbb{R}^+$ and E$_{7(7)}$, with representation indices omitted for legibility. The quantities $\eta_{ij}= \eta_{(ij)}$ and $f_{ab}{}^c = f_{[ab]}{}^c$ are the CSO$(p,q,r)$ quadratic form and the $G_3$ structure constants, respectively. The former can be taken, without loss of generality, as
\begin{eqnarray} \label{eq:Charges2bis}
\eta_{ij} = \textrm{diag} \, \big( 1, \stackrel{p}{\ldots} , 1 , -1 , \stackrel{q}{\ldots} , -1 , 0, \stackrel{r}{\ldots} , 0  \big) \; , \qquad p+q+r =5 \; . 
\end{eqnarray}
The latter is either traceless, $f_{ab}{}^b =0$, or traceful, $f_{ab}{}^b \neq 0$, depending on whether $G_3$ is or is not unimodular. In either case, $f_{ab}{}^c$ is subject to the Jacobi identity
\begin{equation} \label{eq:JIG3}
f_{[ab}{}^d \, f_{c]d}{}^e = 0 \; .
\end{equation}
Finally, $g_1$ and $g_2$ in (\ref{eq:ETtrombone}) are non-vanishing coupling constants.


\subsection{Duality frames for $\mathrm{TCSO}(5,0,0;\mathrm{N})$} \label{sec:TCS500NET}


Most of our discussion will hinge around the $\textrm{TCSO} (5,0,0;\mathrm{N})$ gaugings. For this subset of the $\textrm{TCSO} (p,q,r;\mathrm{N})$ family, the DP embedding tensor is again given by (\ref{eq:ETtromboneSplit}), (\ref{eq:ETtrombone}), with now $\eta_{ij} \equiv \delta_{ij}$ in (\ref{eq:Charges2bis}) and still arbitrary $f_{ab}{}^c = f_{[ab]}{}^c$ subject to (\ref{eq:JIG3}). In this duality frame, the $\textrm{TCSO} (5,0,0;\mathrm{N})$ embedding tensor is manifestly $G_3 \times \textrm{SO}(5)$-covariant. We will also find it helpful to describe the same $\textrm{TCSO} (5,0,0;\mathrm{N})$ theory in two other duality frames that we will call the `special lagrangian (SLAG) twisted frame' and the `associative three-cycle (A3C) twisted frame'. Both have already been used in \cite{Pico:2025cmc}, although not with those names. In these alternate frames, the embedding tensor is only explicitly covariant under $G_3 \times \mathrm{SO}(2) \times \mathrm{SO}(3)$ and $G_3 \times \mathrm{SO}(3)_{-} \times \mathrm{SO}(3)_{+}$, with these SO groups defined as the following subgroups of SO(5):
\begin{equation} \label{eq:SO5subgroups}
\textrm{SLAG:}  \;\;
\textrm{SO}(5)  \supset
\textrm{SO}(2) \times \textrm{SO}(3) \; , \qquad
\textrm{A3C:}  \;\;
\textrm{SO}(5) \supset 
\textrm{SO}(4)  \sim 
\textrm{SO}(3)_- \times \textrm{SO}(3)_+ \;   .
\end{equation}
The SO(5) index $i=4, \ldots , 8$ breaks up accordingly in either frame. In the SLAG case, it splits as $i=(\alpha, \bar{a})$, with $\alpha = 4,5$ and $\bar{a} = 6,7,8$. In the A3C case, it splits using gamma matrices, but we will not need to introduce specific notation for this case. 

Both twisted frames are related to the DP frame (\ref{eq:ETtromboneSplit}), (\ref{eq:ETtrombone}) through duality transformations (\ref{eq:TransXSymbol}). The 
E$_{7(7)}$ group elements,
\begin{equation} \label{eq:E7GroupElement}
U_{\uM}{}^{\uN} =e^{\Upsilon_{\uM}{}^{\uN}} \; , 
\end{equation}
involved in either transformation respectively exponentiate from the E$_{7(7)}$ Lie algebra elements \cite{Pico:2025cmc}
\begin{equation} \label{eq:TransMat}
\textrm{SLAG:}  \;\;
\Upsilon \equiv g_2 g_1^{-1} \big( t_{2368} - t_{1378} \big) \; , \quad
\textrm{A3C:}  \;\;
\Upsilon = \tfrac12   \, g_2 g_1^{-1} \big( t_{1738} + t_{6238}  + t_{1356} + t_{2357}   \big)\;  ,
\end{equation}
where we have again omitted representation indices. In these expressions, $g_1$, $g_2$ are the same couplings that appear in  (\ref{eq:ETtrombone}). For later reference, the SLAG embedding tensor takes on the explicit expression \cite{Pico:2025cmc}
{\setlength\arraycolsep{2pt}
\begin{eqnarray} \label{eq:ETSLAGChanged}
\tilde{X}_M &=& (\tilde{X}_{AB} , \tilde{X}^{AB}) 
= (\tilde{X}_{ab} , \tilde{X}_{ai} , \tilde{X}_{ij} , \tilde{X}^{ab} , \tilde{X}^{ai} , \tilde{X}^{ij}) \nonumber \\[4pt]
&=& (\tilde{X}_{ab} , \tilde{X}_{a\beta} , \tilde{X}_{a\bar{b}} , \tilde{X}_{\alpha\beta}, \tilde{X}_{\alpha \bar{b}}  , \tilde{X}_{\bar{a} \bar{b}}  , \tilde{X}^{ab} , \tilde{X}^{a\beta} , \tilde{X}^{a\bar{b}} , \tilde{X}^{\alpha\beta}, \tilde{X}^{\alpha \bar{b}}  , \tilde{X}^{\bar{a} \bar{b}} ) \; ,
\end{eqnarray}
}with electric components, descending from $\tilde{X}_{AB}$, given by 
{\setlength\arraycolsep{2pt}
\begin{eqnarray} \label{eq:ETSLAGChangedElec}
\tilde{X}_{ab} &=& 0 \; , \nonumber \\[4pt]
\tilde{X}_{a \beta} &=& {X}_{a \beta}  \; , \nonumber \\[4pt]
\tilde{X}_{a \overline{b}} &=&  {X}_{a \overline{b}} + g_2 \, \epsilon_{\overline{b}}{}^{c\overline{d}} \, t_{3a c \overline{d}}  \; , \nonumber \\[4pt]
\tilde  X_{\alpha \beta} &=& X_{\alpha \beta} - g_2^2 g_1^{-1} \epsilon_{\alpha \beta} \,  t_3{}^8 + \tfrac12 \, g_2^2 g_1^{-1} \epsilon_{\alpha \beta} \, \epsilon^{abc} f_{bc}{}^{d} \epsilon_{d\overline{e}3} \,  t_a{}^{\overline{e}}  \; , \nonumber \\[4pt]
  \tilde X_{\alpha \overline{a}} &=& X_{\alpha \overline{a}} +  g_2^2 g_1^{-1} \epsilon_{\alpha\beta} t_3{}^\beta \delta^8_{\overline{a}} + \tfrac12 g_2 \epsilon_{\underline{a}}{}^{bc} t_{\alpha bc 8}
    -  \tfrac12 g_2  \delta^8_{\underline{a}}\epsilon^{bc\overline{d}} t_{\alpha bc \overline{d}} + \tfrac12  g_2^2 g_1^{-1}  \epsilon^{abc} f_{bc}{}^d \epsilon_{\overline{a} d 3} \epsilon_{\alpha \beta} t_{a}^{\beta} \; , \nonumber \\[4pt]
\tilde X_{\overline{a} \overline{b}} &=&  X_{\overline{a} \overline{b}} -  g_2 \, \epsilon_{[\overline{a}}{}^{cd}t_{\overline{b}]bc8} +  g_2  \, \delta^8_{[\overline{a}} \epsilon^{c d \overline{e}} t_{\overline{b}]c d \overline{e}} \; ,
\end{eqnarray}
}and magnetic components, coming from $\tilde{X}^{AB}$,
{\setlength\arraycolsep{1pt}
\begin{eqnarray} \label{eq:ETSLAGChangedMag}
\tilde{X}^{ab} &=& X^{ab} + \tfrac12 g_2^2 g_1^{-1}  \epsilon^{[a|cd} f_{cd}{}^{\overline{e}} \epsilon^{|b]fg} t_{\overline{e}fg8} +g_2^2 g_1^{-1} \epsilon^{8 \overline{e} \overline{f} }\epsilon^{3c[a} \delta^{b]d} t_{cd \overline{e} \overline{f}}  
+ g_2  \epsilon^{ab \overline{c}} t_{\overline{c}}{}^8 -  g_2  \epsilon^{ab}{}_{ \overline{c}} t_8{}^{\overline{c}} \; , \nonumber \\[6pt]
\tilde{X}^{a \beta} &=& {X}^{a \beta}  \; , \nonumber \\[4pt]
\tilde{X}^{a \overline{b}} &=&  {X}^{a\overline{b}} - \tfrac14  g_2^2 g_1^{-1}  \epsilon^{acd}f_{cd}{}^{e} \epsilon^{\overline{b}fg} t_{efg8}- \tfrac14 g_2^2 g_1^{-1} \epsilon^{acd}f_{cd}{}^{\overline{e}} \epsilon^{\overline{b}fg} t_{\overline{e}fg3} \nonumber \\[2pt]
&& + g_2^2 g_1^{-1}   \epsilon^{\overline{b}c3} \epsilon^{d \overline{e} 3}  \delta^{af} t_{fcd \overline{e}} + g_2 \epsilon^{ace} \epsilon^{\overline{b}}{}_{\overline{d}f} \epsilon_{e}{}^{f3} t_c{}^{\overline{d}}  \; , \nonumber \\[6pt]
\tilde X^{\alpha \beta} &=&    \tilde X^{\alpha \overline{b}} = \tilde  X^{\overline{a} \overline{b}} =0   \; .
\end{eqnarray}
}The untilded components, $X_{a \beta}$, etc., that appear on the r.h.s.~of (\ref{eq:ETSLAGChangedElec}), (\ref{eq:ETSLAGChangedMag}) correspond to the original expressions (\ref{eq:ETtrombone}) with $\eta_{ij}= \delta_{ij}$ and branched-out indices under the left-most entry in (\ref{eq:SO5subgroups}), as in (\ref{eq:ETSLAGChanged}).


\section{Eleven-dimensional origin}
\label{sec:CTFrom11D}


In this section, we will show that $D=4$ $\cN=8$ TCSO$(5,0,0;\mathrm{N})$-gauged supergravity locally arises by consistent truncation of $D=11$ supergravity on $G_3 \rtimes S^4$, where the fibrations of $S^4$ on $G_3$ are as in the M5-brane solutions of \cite{Acharya:2000mu,Gauntlett:2006ux}. Prior to doing that, we will review some aspects of the $\Ex{7}$ ExGG formulation of $D=11$ supergravity in order to fix our conventions, and we will construct an auxiliary truncation on the direct product manifold $G_3 \times S^4$.


\subsection{$\Ex{7}$ exceptional generalised geometry for M-theory}
\label{sec:Frames_EGGintro}


Recall that the bosonic field content of $D=11$ supergravity \cite{Cremmer:1978km} contains the metric, $g_{11}$, along with a three-form,  $A$, and a six-form potential, $\tilde{A}$. The forms are on-shell related by Hodge duality in order to ensure the propagation of 128 bosonic degrees of freedom, like in the situation reviewed in section \ref{sec:FieldContentET}. The duality covariant approaches \cite{Coimbra:2011nw,Coimbra:2011ky,Coimbra:2012af,Berman:2010is,Hohm:2013pua,Hohm:2013uia} to $D=11$ supergravity manifestly exhibit the $\mathbb{R}^+ \times \mathrm{E}_{7(7)}$ symmetry of $D=4$ $\cN=8$ ungauged supergravity directly in the higher-dimension. They do so at the expense of sacrificing manifest local GL$(11)$ diffeomorphism covariance in favour of only explicit $\GLg{4}\times \GLg{7}$ covariance. For that reason, this approach is perfectly suited to describe $D=11$ supergravity on a direct product background $M_4 \times M_7$, with $M_4$ and $M_7$ respectively equipped with pseudo-Riemannian and Riemannian metrics.

This product structure is implemented by breaking up the $D=11$ spacetime coordinates as $(x^\mu , y^m)$, with $\mu =0,1,2,3$ as in section \ref{sec:FieldContentET}, and $m=1, \ldots , 7$. The $D=11$ supergravity fields similarly split into $\GLg{4}\times \GLg{7}$-covariant fields as 
\begin{equation} \label{eq:11DFields}
\begin{split}
g_{11}&= e^{2\tilde \Delta} g_{\mu \nu} d x^\mu d x^\nu + g_{mn} Dy^m Dy^n\,,\\[4pt]
A&= \tfrac{1}{3!} A_{mnp} Dy^m \wedge Dy^n \wedge Dy^p + \tfrac{1}{2!} A_{\mu np} \dd x^\mu \wedge Dy^n \wedge Dy^p + \dots\,,
\\[4pt]
\tilde A &= \tfrac{1}{6!} \tilde A_{m_1 \dots m_6} Dy^{m_1} \wedge \dots \wedge Dy^{m_6} + \tfrac{1}{5!}  \tilde A_{\mu m_1 \dots m_5} d x^\mu \wedge Dy^{m_1} \wedge \dots \wedge Dy^{m_5} + \dots\,,
\end{split}
\end{equation}
where $Dy^m \equiv d y^m + A_\mu{}^m dx^\mu$. All fields in (\ref{eq:11DFields}) depend on all $D=11$ coordinates,~{\it i.e.}~$g_{\mu\nu} = g_{\mu\nu} (x,y)$,~etc.,~and the dots stand for higher-order forms of GL$(4)$ whose explicit expression will not be needed. Now, the $\GLg{4}\times \GLg{7}$-covariant fields in (\ref{eq:11DFields}) can be equivalently regarded as components of the following manifestly $\GLg{4}\times  \mathbb{R}^+ \times \mathrm{SU}(8)$ or $\GLg{4} \times \mathbb{R}^+ \times \mathrm{E}_{7(7)}$ covariant objects
\begin{eqnarray} \label{eq:ExObjects}
(G^{-1})^{MN} \Leftrightarrow \big( \tilde \Delta , g_{mn},A_{mnp}, \tilde{A}_{m_1 \dots m_6} \big) \; , \quad 
\cA_\mu{}^M = \big( A_\mu{}^m, A_{\mu mn},  A_{\mu m n p q r} ,  A_{\mu \, m; n_1 \dots n_7} \big) \, , \;
\end{eqnarray}
along with similar objects obtained out of the fields hidden under the ellipses in (\ref{eq:11DFields}). In particular, the $\GLg{4}$ one-form $\cA_\mu{}^M$ sits in the fundamental representation of $\mathrm{E}_{7(7)}$ (and it is only schematic as shown in (\ref{eq:ExObjects}), as further non-linear redefinitions of its components are needed for its supersymmetry variations to comply with the lower-dimensional tensor hierarchy, see \cite{Godazgar:2013dma,Varela:2015ywx}). This can be seen by assigning the four blocks that make up $\cA_\mu{}^M$ in (\ref{eq:ExObjects}) with each of the four $\GLg{7}$ representations that appear in the branching 
\begin{equation} \label{eq:56Branching}
\bm{56} \rightarrow \bm{7}^\prime + \bm{21} + \bm{21}^\prime + \bm{7} 
\end{equation}
under $\mathbb{R}^+ \times \mathrm{E}_{7(7)} \supset \GLg{7}$, with $\mathbb{R}^+ $ charges omitted. The $\bm{7}$ fields $A_{\mu \, m; n_1 \dots n_7}$ are related to the dual graviton and do not feature explicitly in the decomposition (\ref{eq:11DFields}). 

Similarly, $(G^{-1})^{MN}$ is a $\GLg{4}$-invariant (inverse) metric on $\mathbb{R}^+ \times \mathrm{E}_{7(7)}/\mathrm{SU}(8)$. Accordingly, a generalised vielbein $E_M{}^{\uN}$, with inverse $\hat{E}^{M}{}_{\uN}$, can be defined such that 
\begin{eqnarray} \label{eq:GenMet}
G_{MN} (x,y)= \delta_{\uP\uQ} E_M{}^{\uP} (x,y) E_N{}^{\uQ} (x,y) \; , \;
(G^{-1})^{MN} (x,y)= \delta^{\uP\uQ} \hat{E}^M{}_{\uP} (x,y) \hat{E}^N{}_{\uQ} (x,y) . \;
\end{eqnarray} 
Indices $M$ and $\uM$ here and elsewhere are E$_{7(7)}$ curved and flat indices, respectively. In (\ref{eq:GenMet}) we have reinstated the $(x,y)$ dependence of the fields to emphasise that these are all $D=11$ objects. For later reference, the explicit correspondence between some relevant components of the generalised metric $(G^{-1})^{MN}$ and its $\GLg{4}\times \GLg{7}$-covariant building blocks (\ref{eq:ExObjects}) is
\begin{equation} \label{eq:Frames_UpliftExpressions}
\begin{split}
(G^{-1})^{m,n}&= e^{2 \tilde \Delta} g^{mn}\,,\\[4pt]
(G^{-1})^{m}{}_{np}&= e^{2 \tilde \Delta} g^{mq}A_{q n p}\,,\\[4pt]
(G^{-1})^{m,np}&= e^{2 \tilde \Delta} g^{mq}(5 A_{q[n_1 n_2}A_{n_3 n_4 n_5]}+ \tilde A_{q n_1 n_2 n_3 n_4 n_5}) \tfrac{1}{5!}\epsilon^{npn_1 n_2 n_3 n_4 n_5}\,,
\end{split}
\end{equation}
along with
\begin{equation} \label{eq:GDet}
(\det G_{MN})^{- \tfrac{1}{56}}= (\det E^M{}_{A})^{ \tfrac{1}{ 28}}=\sqrt{ \det g_{mn}} \, e^{2 \tilde \Delta} \; .
\end{equation}
For the specific components given therein, (\ref{eq:Frames_UpliftExpressions}) makes precise the relation denoted schematically with the double equivalence symbol in (\ref{eq:ExObjects}). 

The ExGG fields $G_{MN}(x,y)$, $\cA_\mu{}^M (x,y)$, etc., are covariant under E$_{7(7)}$ generalised diffeomorphisms, namely, under the combined action of GL(7) diffeomorphisms and gauge transformations of the component supergravity fields. The action of generalised diffeomorphisms is implemented with the help of the generalised Lie derivative, $L$. For any two generalised vectors $V^M$, $W^M$, this is defined as \cite{Coimbra:2011ky}
\begin{equation} \label{eq:LieDer}
L_V W^M \equiv   V^N \partial_N W^M - (\partial \times_{\text{ad}}V)_N{}^M W^N\, .
\end{equation}
Here, $\partial_M = (\partial_m,0,0,0)$ is defined following the branching (\ref{eq:56Branching}), and we have used the notation, valid for a generalised vector $V^M$ and co-vector $Z_M$, 
\begin{equation} \label{eq:adtimes}
(V \times_{\text{ad}} Z )_N{}^M \equiv (\mathbb{P}_{ \mathbb{R}^+ \times \Ex{7} })^M{}_N{}^K{}_L V^L Z_K \; , 
\end{equation}
where 
$
(\mathbb{P}_{ \mathbb{R}^+ \times \Ex{7} })^M{}_N{}^K{}_L = 12 \, \mathbb{P}^M{}_N{}^K{}_L - \tfrac12 \, \delta^M_N \delta^K_L\
$
is the projector to the adjoint of $ \mathbb{R}^+\times \Ex{7}$, expressed in terms of the projector $\mathbb{P}^M{}_N{}^K{}_L=(t_\alpha)^M{}_N(t^\alpha)^K{}_L$ to the adjoint of $\Ex{7}$.

In ExGG, the E$_{7(7)}$-covariant fields can be regarded as sections of appropriate fibre bundles that generalise the usual tangent bundle of ordinary geometry. Thus, this language is amenable to the introduction of generalised $G$-structures, namely, reductions to bundles with (compact) structure group $G \subset \textrm{SU}(8) \subset \textrm{E}_{7(7)}$. Generalised $G$-structures are completely characterised by $G$-invariant tensors along with an intrinsic torsion in the representations of $G$ that branch from the $\bm{56}_{-1} + \bm{912}_{-1}$ of $\mathbb{R}^+ \times \mathrm{E}_{7(7)}$ under $G \subset \textrm{SU}(8) \subset \textrm{E}_{7(7)}$. In particular, the generalised metric (\ref{eq:GenMet}) defines a generalised SU$(8)$-structure with intrinsic torsion $X_{\uM \uN}{}^{\uP} (x,y)$ determined through the generalised Lie derivative (\ref{eq:LieDer}) as
\begin{equation} \label{eq:IntTor}
L_{\hat E_{\uM} (x,y)} \hat E_{\uN}(x,y) = - X_{\uM\uN}{}^{\uP} (x,y) \, \hat E_{\uP} (x,y)  \; .
\end{equation}
Here, we have again written the $(x,y)$ dependence explicitly as a reminder that all these quantities are eleven-dimensional.


\subsection{Consistent $\cN=8$ truncation on $G_3 \times S^4$} 

\label{sec:TruncDP}


As we have just emphasised, the expressions (\ref{eq:GenMet}) and (\ref{eq:IntTor}) that respectively define the (inverse) generalised frame $\hat{E}^M{}_{\uN} (x,y)$ and its associated intrinsic torsion $X_{\uM\uN}{}^{\uP} (x,y)$ are  completely general in ExGG. Now, in this language, a consistent truncation of $D=11$ supergravity on a seven-dimensional manifold $M_7$ down to $D=4$ supergravity exists if the SU$(8)$-structure specified by the generalised metric is reduced to a (local) $G$-structure with E$_{7(7)}$-singlet, constant  intrinsic torsion \cite{Lee:2014mla,Cassani:2019vcl}. In particular, a constant-torsion GIS, $G=\bm{1}$, leads to a maximally supersymmetric  consistent truncation \cite{Lee:2014mla}. Thus, in practical terms, a consistent truncation of $D=11$ supergravity down to $D=4$ $\cN=8$ supergravity exists if the ExGG fields (\ref{eq:ExObjects}) factorise in terms of the four-dimensional fields, $M^{\uM\uN}$, $A^{\uM}$, etc. reviewed in section \ref{sec:FieldContentET} through the generalised Scherk-Schwarz expressions \cite{Berman:2012uy,Lee:2014mla}
\begin{eqnarray} \label{eq:GenSS}
& \tilde \Delta(x,y) = \tilde \Delta(y) \; , \qquad
g_{\mu \nu}(x,y)=  g_{\mu \nu}(x) \; , \nonumber \\[4pt]
& (G^{-1})^{MN} (x,y) =M^{\uP\uQ}(x) \, \hat{E}^M{}_{\uP}(y) \hat{E}^N{}_{\uQ}(y) \; , \quad 
\cA_\mu{}^M (x,y) = \hat{E}^M{}_{\uN}(y) \, A_\mu{}^{\uN}(x) \; , \;
\end{eqnarray}
and (\ref{eq:IntTor}) holds with constant intrinsic torsion $X_{\uM\uN}{}^{\uP}$  \cite{Lee:2014mla,Cassani:2019vcl},
\begin{equation} \label{eq:CTCon}
L_{\hat E_{\uM} (y)} \hat E_{\uN}(y) = - X_{\uM\uN}{}^{\uP} \, \hat E_{\uP} (y)  \; .
\end{equation}
The constant instrinsic torsion $X_{\uM\uN}{}^{\uP}$ then defines the embedding tensor of the $D=4$ $\cN=8$ supergravity \cite{Berman:2012uy,Lee:2014mla}.

Equipped with these prescriptions we will show, in section \ref{sec:TruncTwisted}, that $D=11$ supergravity admits a maximally supersymmetric consistent truncation on the twisted internal manifold $\Sigma_3 \rtimes S^4$ corresponding to the AdS$_4$ near horizon geometries associated to the wrapped M5-brane solutions of \cite{Acharya:2000mu,Gauntlett:2006ux}. In this section, we will start by showing the existence of a related $D=11$ truncation on the direct product manifold $G_3 \times S^4$, with group $G_3$ of Bianchi type N, down to $D=4$ $\cN=8$ TCSO$(5,0,0;\mathrm{N})$-gauged supergravity as formulated in section \ref{sec:FieldContentET}. By the previous discussion, in order to do this, we must construct a constant intrinsic torsion GIS on $G_3 \times S^4$. In other words, we must build an (inverse) generalised frame $\hat{E}^M{}_{\uN}(y)$ out of geometric and background supergravity data on $G_3 \times S^4$, and then verify that (\ref{eq:CTCon}) holds with $X_{\uM}$ identified with the DP embedding tensor (\ref{eq:ETtromboneSplit}), (\ref{eq:ETtrombone}). As we will see momentarily, the basic building blocks for the GIS on $G_3 \times S^4$ will be the conventional parallelisation on the  group manifold $G_3$ and the generalised parallelisation on the $S^4$ given in \cite{Lee:2014mla}.

In order to construct our GIS, it is helpful to regard the inverse generalised frame $\hat{E}^M{}_{\uN}(y)$ as a collection, labelled by the flat index $\uN$, of $\bm{56}$ generalised vectors, each of them with generalised vector index $M$. At fixed $\uN$, each of these generalised vectors splits, by (\ref{eq:56Branching}), into $\textrm{GL}(7)$-covariant components as in the rightmost relation in (\ref{eq:ExObjects}), namely, 
\begin{equation} \label{eq:GenFrameSplit}
\hat{E}^M{}_{\uN}  = \big( \hat{E}^m{}_{\uN}  , \hat{E}_{mn \, \uN}  ,  \hat{E}_{mnpqr \, \uN}  ,  \hat{E}_{m; n_1 \dots n_7 \, \uN}  \big) \; .
\end{equation}
To define each of these components, it is helpful to also split the flat index $\uN$ under, this time, $\mathbb{R}^+ \times  \textrm{E}_{7(7)} \supset \textrm{GL}(3) \times \textrm{SL}(5)$ instead, because the GL(7)-covariant quantities on $G_3 \times S^4$ that we will use come in representations of that group. Under $\mathbb{R}^+ \times  \textrm{E}_{7(7)} \supset \textrm{GL}(3) \times \textrm{SL}(5)$, and omitting $\mathbb{R}^+$ charges,
\begin{equation} \label{eq:SplitGL3SL5}
\bm{56} \rightarrow (\bm{3} , \bm{1}) + (\bm{3} , \bm{5}) + (\bm{1} , \bm{10}) + (\bm{3}^\prime , \bm{1}) + (\bm{3}^\prime , \bm{5}^\prime)  + (\bm{1} , \bm{10}^\prime) \; ,
\end{equation}
so that, at fixed value of the curved index $M$, the flat index $\uN$ accordingly branches as
\begin{equation} \label{eq:GenFrameSplit2}
\hat{E}^M{}_{\uN}  = \big(\epsilon_{abc}\hat E^{M \, c} \; , \;  - \tfrac12 \epsilon_{abc} \hat E^{M \, bc}{}_i \; , \;  \hat E^M{}_{ij} \; , \;  \epsilon^{abc}\hat E'^M{}_c \; , \;  - \hat E'^{M \,ia} \; , \;  \hat E'^{M \,ij}   \big) \; ,
\end{equation}
with indices $a=1,2,3$ and $i = 4, \ldots 8$ ranging as in section~\ref{sec:FieldContentET}. 

Finally, we propose the following expressions for the various blocks in (\ref{eq:GenFrameSplit2}):
{\setlength\arraycolsep{2pt}
\begin{eqnarray} \label{eq:GenFrameG3S4}
\hat E^a &=& \big( 0 \; , \; 0 \; , \; 0 \; , \; R_3\,  e^a \otimes (\text{vol}_{4} \wedge \text{vol}_{3}) \big) \; , \nonumber \\[5pt]
\hat E^{ab}{}_i \;  & =&   \big( 0 \; , \;  R_3^2 \, y_i \, e^a \wedge e^b \; , \; R_3^2 \, e^a \wedge e^b \wedge \left( R *_4 \, d y_i +y_i \, A \right) \; , \; 0 \big) \; , \nonumber \\[5pt]
\hat E_{ij} \;  &=& \big( v_{ij} \; , \;  R^2 *_4 ( d y_i \wedge d y_j) + \imath_{v_{ij}} A \; , \;  0 \; , \;  0 \big) \; , \nonumber \\[5pt]
 \hat E'_a  \; & =& \big( R_3^{-1}\hat e_a \; , \;  0 \; , \;  0 \; , \;  0 \big) \; ,  \\[5pt]
\hat E'^{ia} \;  &=& \big( 0 \; , \; R \, R_3 \, e^a \wedge d y^i  \; , \; R_3\, e^a \wedge \left( R \, d y^i\wedge A - y^i \, \text{vol}_{4} \right) \; , \; 0  \big) \; , \nonumber \\[5pt]
\hat E'^{ij}&=&  \big( 0 \; , \; 0 \; , \; R^2 (d y^i \wedge d y^j)\wedge \text{vol}_{3} \; , \;  R^2 \, jA \wedge  ( d y^i \wedge d y^j)\wedge \text{vol}_{3} -  v^{ij\,\flat} \otimes (\text{vol}_{4} \wedge \text{vol}_{3}) \big) \; , \nonumber
\end{eqnarray}
}with the curved index $M$ omitted on the l.h.s.'s, and split on the r.h.s.'s as in (\ref{eq:GenFrameSplit}). Here, $e^a$ is a left-, say, $G_3$-invariant vielbein on the $G_3$ group manifold, with inverse $\hat{e}_a$, and $\vol_3$ is the volume form on $G_3$. Also, $y^i$ are constrained coordinates on $\mathbb{R}^5$ that define $S^4$ as the locus (\ref{eq:SdinRd+1}), $v^{ij}$ are the SO$(5)$ Killing vectors of $S^4$, and $A$ is a local three-form potential for the volume on $S^4$, $dA = 3 R^{-1} \, \vol_4$. The Hodge dual $*_4$ is taken w.r.t.~the round metric on $S^4$, $\imath_{v}$ denotes the interior product w.r.t.~a conventional vector $v$, we use the operator $j$ as defined in \cite{Pacheco:2008ps, Coimbra:2011ky}, and $v^{\flat}$ stands for the one-form dual to the geometric vector $v$. Finally, the constants $R_3$ and $R$ respectively set scales on $G_3$ and $S^4$, {\it i.e.}, $R$ is the $S^4$ radius. Please see appendix~\ref{sec:Conventions} for our conventions on these operations and quantities.

As already advertised, the generalised frame (\ref{eq:GenFrameSplit2}),  (\ref{eq:GenFrameG3S4}) builds on the respective ordinary \cite{Scherk:1979zr} and generalised \cite{Lee:2014mla} parallelisations of $G_3$ and $S^4$, and uses various generalised bundles on $S^4$ introduced in \cite{Lee:2014mla,Cassani:2019vcl}. Using (\ref{eq:LieDer}), a long computation reveals that the generalised Lie derivative of the frames (\ref{eq:GenFrameSplit2}), (\ref{eq:GenFrameG3S4}) do indeed obey the GIS condition (\ref{eq:CTCon})
with constant intrinsic torsion $X_{\uM \uN}{}^{\uP}$ given, precisely, by the DP embedding tensor (\ref{eq:ETtromboneSplit}), (\ref{eq:ETtrombone}) with the coupling constants and radii identifications 
\begin{equation} \label{eq:CouplingsRadii}
g_1=R^{-1} \; , \qquad 
g_2=R_3^{-1} \; .
\end{equation}
The details can be found in appendix~\ref{sec:GISDetails}. This establishes the consistency of the truncation of $D=11$ supergravity on $G_3 \times S^4$ down to $D=4$ $\cN=8$ TCSO$(5,0,0;\mathrm{N})$-gauged supergravity. 

The explicit embedding of the $D=4$ $\cN=8$ fields into the $D=11$ fields of the conventional formulation of \cite{Cremmer:1978km} can be obtained by bringing the frame (\ref{eq:GenFrameSplit2}),  (\ref{eq:GenFrameG3S4}) to the generalised Scherk-Schwarz expressions (\ref{eq:GenSS}), and then unpacking the ExGG fields using (\ref{eq:ExObjects}), (\ref{eq:Frames_UpliftExpressions}). For example, for the internal components of the (inverse) $D=11$ metric along $G_3 \times S^4$ we obtain, after some calculation,
{\setlength\arraycolsep{-.5pt}
\begin{eqnarray} \label{eq:DPUpliftedMetric}
&& e^{2 \tilde \Delta} g^{mn} =  \tfrac14 R_3^{-2} \epsilon^{acd} \epsilon^{bef} M_{cd \, ef} \, \hat e^{m}{}_a \, \hat e^{n}{}_b + \tfrac14  M^{ij \, kl} \, v^{m}{}_{ij} \, v^{n}{}_{kl} + \tfrac12 R_3^{-1} \epsilon^{abc} M_{bc}{}^{ij} \, \hat e^{(m}{}_a \, v^{n)}{}_{ij} . \qquad \;
\end{eqnarray}
}Here, $M_{cd \, ef}$, $M^{ij \, kl}$ and $M_{bc}{}^{ij}$ are some of the $\textrm{GL}(3) \times \textrm{SL}(5)$-covariant components of the $D=4$ scalar matrix $M^{\uM \uN}$ that result from the splitting (\ref{eq:SplitGL3SL5}). We have also reintroduced curved GL$(7)$ indices $m=1, \ldots , 7$, with the understanding that, on $\hat{e}^{m}{}_a$ and $v^{m}{}_{ij}$, these only range non-trivially along the $G_3$ and $S^4$ directions, respectively.

The expression (\ref{eq:DPUpliftedMetric}) makes manifest that our $D=11$  truncation on $G_3 \times S^4$ can indeed be understood as a combination of conventional, on $G_3$ \cite{Scherk:1979zr}, and generalised, on $S^4$ \cite{Lee:2014mla}, Scherk-Schwarz reductions. For vanishing $D=4$ scalars, the various blocks of $M^{\uM \uN}$ reduce to $M_{cd \, ef} = 2 \delta_{e[c} \delta_{d]f}$, $M^{ij \, kl} = 2 \delta^{k[i} \delta^{j]l}$ and $M_{bc}{}^{ij}=0$, and (\ref{eq:DPUpliftedMetric}) reduces, upon inversion,  to the standard $G_3 \times \textrm{SO}(5)$-invariant (or $\textrm{SU}(2)_{\textrm{left}} \times \textrm{SU}(2)_{\textrm{right}} \times \textrm{SO}(5)$-invariant, if $G_3=\textrm{SU}(2)$) metric on $G_3 \times S^4$,
\begin{equation} \label{eq:DP7dMetric}
g_{7} = R_3^2 \, \delta_{ab} \, e^a e^b + R^2  \delta_{ij} \,   dy^i dy^j \; .
\end{equation}
%



\subsection{Consistent $\cN=8$ truncation on $G_3 \rtimes S^4$} 

\label{sec:TruncTwisted}


We will now build on the results of section~\ref{sec:TruncDP} to show that $D=4$ $\cN=8$ TCSO$(5,0,0;\mathrm{N})$-gauged supergravity also arises locally by consistent truncation of $D=11$ supergravity on the twisted manifold $G_3 \rtimes S^4$. We will consider two possible fibrations of $S^4$ over $G_3$, corresponding to the two different supersymmetric solutions of \cite{Acharya:2000mu,Gauntlett:2006ux}. 

The near-horizon M5-brane solutions $\textrm{AdS}_4 \times (\Sigma_3 \rtimes S^4)$ of \cite{Acharya:2000mu,Gauntlett:2006ux} implement supersymmetry via topological twist \cite{Witten:1988ze,Bershadsky:1995qy}, with the fibration governed by the spin connection of $\Sigma_3$. In section~\ref{sec:Subsectors} we will recover and write explicitly the relevant background metrics on $\Sigma_3 \rtimes S^4$ for the two different solutions of \cite{Acharya:2000mu,Gauntlett:2006ux}. For now, we begin by replacing $\Sigma_3$ with the group manifold $G_3$, as in section~\ref{sec:TruncDP}, and consider the spin connection one-form $\omega^{ab}$ of $G_3$. This is defined, as usual, by $de^a = -\omega^{ab} \wedge e_b$, with $e^a$ the left-invariant vielbein on $G_3$. Geometrically, the effect of the twist \cite{Witten:1988ze,Bershadsky:1995qy} in the background metric $g_7$ on $G_3 \rtimes S^4$ is to replace the ordinary derivatives $dy^i$ in (\ref{eq:DP7dMetric}) with covariant derivatives $D y^i$ governed by $\omega^{ab}$. There are two possible ways to introduce such covariant derivatives, corresponding to the two different branchings (\ref{eq:SO5subgroups}) of SO(5) to SO(3) subgroups thereof. We thus have
\begin{eqnarray} \label{eq:CovDers}
\textrm{SLAG:} \,
Dy^\alpha = dy^\alpha  , \, 
Dy^{\bar{a}} = dy^{\bar{a}} + \delta^{\bar{a}}_a \, \delta^b_{\bar{b}} \, \omega^a{}_b \, y^{\bar b} \; , \;\;
\textrm{A3C:} \,
Dy^i = dy^i -\tfrac14 \epsilon_{abc} \, \omega^{ab} \, (J^c_-)^i{}_j \, y^j . \quad
\end{eqnarray}
In the SLAG case, $i=(\alpha, \bar{a})$, with $\alpha = 4,5$ and $\bar{a} = 6,7,8$ labelling the fundamental representations of the SO(2) and SO(3) that appear in the leftmost relation in (\ref{eq:SO5subgroups}). In the A3C case, $(J^c_\pm)^i{}_j$ are the generators of $\textrm{SO}(3)_\pm$ in the $(\bm{2},\bm{2}) + (\bm{1},\bm{1})$ representation.

To show the consistency of the truncation of $D=11$ supergravity on either twisted manifold $G_3 \rtimes S^4$ we must show the existence of respective GISs on $G_3 \rtimes S^4$ with constant intrinsic torsion. Let us construct these GISs starting from the GIS built in section~\ref{sec:TruncDP} on the direct product manifold $G_3 \times S^4$. To do this, we will adapt a prescription given in \cite{Cassani:2019vcl,Cassani:2020cod}, and also used in a similar context in~\cite{Bhattacharya:2024tjw}, to construct these type of twists in ExGG. More concretely, the twisted generalised (inverse) frame, $\hat{\tilde{E}}^M{}_{\uN} (y)$, on the fibred manifold $G_3 \rtimes S^4$ should be obtained from its direct product counterpart, $\hat{E}^M{}_{\uN} (y)$ in (\ref{eq:GenFrameSplit2}), (\ref{eq:GenFrameG3S4}), by the action of a suitable $\mathbb{R}^+ \times \mathrm{E}_{7(7)}$ transformation $U(y)$ taking values on $G_3 \times S^4$:
\begin{equation} \label{eq:TwistedFrame}
\hat{\tilde{E}}^M{}_{\uP} (y)= U_N{}^M (y) \, \hat{E}^N{}_{\uP} (y) \; .
\end{equation}
The twists considered in \cite{Cassani:2019vcl,Cassani:2020cod,Bhattacharya:2024tjw} are governed by an abelian spin connection $\upsilon$. Their prescription entails promoting $\upsilon$ to a generalised covector, and then tensoring it with a generalised vector containing the specific Killing vector on $S^4$ gauged by the fibration. This tensor product is then projected with (\ref{eq:adtimes}) back into the Lie algebra of $\mathbb{R}^+ \times \mathrm{E}_{7(7)}$, and finally exponentiated to obtain the finite $\mathbb{R}^+ \times \mathrm{E}_{7(7)}$ transformation $U(y)$.

Let us now replicate those steps in our present non-abelian case. Firstly, we start by promoting the spin connection $\omega_m{}^{ab}$ of $G_3$ to a generalised covector $\omega_M{}^{ab} = (\omega_m{}^{ab},0,0,0)$ with the help of (\ref{eq:56Branching}). Again, we have reintroduced GL$(7)$-covariant curved indices $m=1, \ldots , 7$ on $G_3 \times S^4$ with the understanding that $\omega_m{}^{ab}$ takes non-trivial values only along $G_3$. Secondly, we should identify, among the $\bm{56}$ generalised vectors $\hat{E}^M{}_{\uN} (y)$ making up the frame on $G_3 \times S^4$, those that contain the $S^4$ Killing vectors involved in either twisting (\ref{eq:CovDers}). By inspection of (\ref{eq:GenFrameG3S4}), we see that only $\hat{E}_{ij}$ contains in its GL$(7)$-vector component, the first in the decomposition (\ref{eq:GenFrameSplit}), the Killing vectors $v_{ij}$ of $S^4$. Thus, we should further break the SO(5) indices $ij$ under (\ref{eq:SO5subgroups}) in either case, and further select only those generalised vectors within $\hat{E}_{ij}$ that contain the specific $S^4$ Killing vectors involved in the twistings (\ref{eq:CovDers}). These are $\hat{E}_{\bar{a}\bar{b}}$ in the SLAG case, and $(J_-^c)^{ij} \hat{E}_{ij}$ in the A3C case. Next, we construct the $G_3 \times S^4$-valued $\mathbb{R}^+ \times \textrm{E}_{7(7)}$ Lie algebra elements $\Upsilon(y)$ given by
\begin{equation} \label{eq:ExGGTwist}
\textrm{SLAG:} \; 
\Upsilon_M{}^N= \tfrac{R}{R_3} \, \delta_a^{\bar a} \delta_b^{\bar b}\, \big( \omega^{ab} \times_\text{ad}  \hat{E}_{\bar a \bar b} \big)_M{}^N  , \quad 
\textrm{A3C:} \;
\Upsilon_M{}^N=-\tfrac14 \tfrac{R}{R_3} \, \epsilon_{abc} (J_-^c)^{ij} \big( \omega^{ab} \times_\text{ad}  \hat{E}_{ij} \big)_M{}^N , 
\end{equation}
in either case. Here, we have restored curved E$_{7(7)}$ indices and have used the definition (\ref{eq:adtimes}). Lastly, for either Lie algebra element (\ref{eq:ExGGTwist}), the corresponding $G_3 \times S^4$-valued finite E$_{7(7)}$ group element is
\begin{equation} \label{eq:E7GroupElement}
U_{M}{}^{N} (y) =e^{\Upsilon_{M}{}^{N}(y)} \; .
\end{equation}

The generalised frame $\hat{\tilde{E}}^M{}_{\uP} (y)$ on either fibred manifold $G_3 \rtimes S^4$, SLAG or A3C, is thus finally given by (\ref{eq:TwistedFrame}) with the direct product frame $\hat{E}^M{}_{\uP} (y)$ in (\ref{eq:GenFrameSplit2}), (\ref{eq:GenFrameG3S4}), and the transformation $U_{M}{}^{N} (y)$ in (\ref{eq:E7GroupElement}) that exponentiates from either $\mathbb{R}^+ \times \mathrm{E}_{7(7)}$ Lie algebra element $\Upsilon_{M}{}^{N} (y)$ in (\ref{eq:ExGGTwist}). We have computed the generalised metric (\ref{eq:GenMet}) on $G_3 \rtimes S^4$ associated to $\hat{\tilde{E}}^M{}_{\uP} (y)$, and have extracted the components of the geometric metric $g_{mn}$ using (\ref{eq:Frames_UpliftExpressions}). As expected, the resulting metric does indeed involve the covariant derivatives (\ref{eq:CovDers}) responsible for the fibration of $S^4$ over $G_3$.

One can finally ask whether the generalised frames $\hat{\tilde{E}}^M{}_{\uP} (y)$ that we have just described for either fibred manifold $G_3 \rtimes S^4$, SLAG or A3C, lead to consistent truncations of $D=11$ supergravity down to $D=4$ $\cN=8$ gauged supergravity. The answer is positive in both cases, as long calculations corroborate. In the SLAG case, the twisted frame $\hat{\tilde{E}}^M{}_{\uP} (y)$ does indeed satisfy the GIS condition (\ref{eq:CTCon}) with constant intrinsic torsion given by the $D=4$ $\cN=8$ embedding tensor (\ref{eq:ETSLAGChanged}) with (\ref{eq:ETSLAGChangedElec}), (\ref{eq:ETSLAGChangedMag}). Similarly, the A3C twisted frame $\hat{\tilde{E}}^M{}_{\uP} (y)$ does also satisfy the GIS condition (\ref{eq:CTCon}) with constant intrinsic torsion given by the $D=4$ $\cN=8$ embedding tensor specified in section 4.3 of \cite{Pico:2025cmc}. Since $\Sigma_3 \equiv H^3$ is locally diffeomorphic to the group manifold $G_3$ of Bianchi type V, this analysis shows, in particular, that $D=11$ supergravity admits a local consistent truncation to $D=4$ $\cN=8$ supergravity on the internal geometries $\Sigma_3 \rtimes S^4$ of the supersymmetric wrapped M5-brane solutions of 
\cite{Acharya:2000mu,Gauntlett:2006ux}.


\subsection{Equivalence of the direct product and twisted truncations} 

\label{sec:TruncEquivalence}


As reviewed in section \ref{sec:TCS500NET}, the two $D=4$ $\cN=8$ gauged supergravities encountered by the twisted consistent truncations of section \ref{sec:TruncTwisted} are related by duality transformations between each other. They are also duality-related to the truncated $D=4$ $\cN=8$ theory obtained in section \ref{sec:TruncDP}. All three truncated theories are therefore equivalent. In conclusion, $D=11$ supergravity admits consistent truncations on the direct product manifold $G_3 \times S^4$, and on the SLAG and A3C fibred manifolds $G_3 \rtimes S^4$, down to one and the same maximal four-dimensional gauged supergravity: the TCSO$(5,0,0;\mathrm{N})$ theory. 

The reason for the equivalence of these $D=11$ truncations on the distinct manifolds $G_3 \times S^4$ and $G_3 \rtimes S^4$ is that the ExGG twisting process described in section \ref{sec:TruncTwisted} following \cite{Cassani:2019vcl,Cassani:2020cod}, and the change of $D=4$ $\cN=8$ duality frames \cite{Pico:2025cmc} reviewed in section \ref{sec:TCS500NET}, are in fact related. Remarkably, the constant, $y$-independent E$_{7(7)}$ Lie algebra elements $\Upsilon_{\uM}{}^{\uN}$ in (\ref{eq:TransMat}) that generate the duality transformation in the $D=4$ supergravity turn out to be the flattened version of the $G_3 \times S^4$-valued E$_{7(7)}$ Lie algebra elements $\Upsilon_{M}{}^{N}(y)$ in (\ref{eq:ExGGTwist}) that turn on the fibration in the ExGG description of the manifold $G_3 \rtimes S^4$. Namely,
\begin{equation} \label{eq:FlatteningCon}
\Upsilon_{\uM}{}^{\uN} = \Upsilon_{M}{}^{N}(y) \, \hat{E}^M{}_{\uM} (y) \, E_N{}^{\uN} (y) \; , 
\end{equation}
where $\hat{E}^M{}_{\uM} (y)$ is the (inverse) generalised frame on $G_3 \times S^4$ given in (\ref{eq:GenFrameSplit2}), (\ref{eq:GenFrameG3S4}), and $E_N{}^{\uN} (y)$ is the corresponding direct frame (see (\ref{eq:DirGenFrameG3S4})).


\section{Consistent subsectors} \label{sec:Subsectors}


We will now particularise the $\cN=8$ truncations discussed in section \ref{sec:CTFrom11D} to specific subsectors that contain the M5-brane AdS$_4$ solutions of \cite{Acharya:2000mu,Gauntlett:2006ux}. We will use the language of generalised $G$-structures to do this, building on \cite{Cassani:2019vcl,Josse:2025uro}. Dealing with a smaller number of $D=4$ fields will allow us to obtain explicit expressions for their embedding into their $D=11$ counterparts in the usual formulation of \cite{Cremmer:1978km}.


\subsection{Ordinary and generalised $\mathrm{SO}(3)$-structures} \label{sec:SO3S}


As discussed in section~\ref{sec:TruncTwisted}, the supersymmetric $\textrm{AdS}_4 \times (\Sigma_3 \rtimes S^4)$ wrapped M5-brane configurations of \cite{Acharya:2000mu,Gauntlett:2006ux} enjoy a constant-torsion GIS in ExGG when $\Sigma_3$ is replaced with a group manifold $G_3$. This GIS is a reduction of the generic generalised SU(8)-structure associated to the generalised ExGG metric $G_{MN}(x,y)$. In this section, we would like to reinterpret these configurations from the point of view of appropriate intermediate generalised SO$(3)_S$-structures, with $\bm{1} \subset \textrm{SO}(3)_S\subset \textrm{SU}(8)$, and $\textrm{SO}(3)_S$ embedded into SU$(8)$ in two different ways, corresponding to the SLAG and A3C cases. By the general prescription of \cite{Cassani:2019vcl}, the background $G_3 \rtimes S^4$ regarded as a generalised SO$(3)_S$-structure manifold must lead to a submaximal consistent truncation. Depending on the embedding of SO$(3)_S$ into SU$(8)$, the resulting $D=4$ supergravity will be $\cN=2$ or $\cN=1$ \cite{Josse:2025uro}. In our case, by construction, these submaximal theories must be consistent subsectors of $D=4$ $\cN=8$ TCSO$(5,0,0;\mathrm{N})$ supergravity. Of course, these will be none other than the SO$(3)_S$-invariant subsectors considered in section 4 of \cite{Pico:2025cmc}. 

In the SLAG and the A3C cases, the supersymmetric M5-brane configurations $\textrm{AdS}_4 \times (\Sigma_3 \rtimes S^4)$ of interest are generically equipped with an ordinary $\textrm{SU}(2)_S \sim \textrm{SO}(3)_S$ and an SU$(3)_S$ structure in conventional geometry, respectively \cite{Gauntlett:2006ux}. One may argue that, for the specific gauged-supergravity-inspired solution of \cite{Acharya:2000mu}, rederived in section 9.4 of \cite{Gauntlett:2006ux}, the A3C structure should be further reduced to the $\textrm{SO}(3)_S^\prime$ real subgroup of SU$(3)_S$. The intuition is that both the SLAG and the A3C cases should be equipped with conventional SO$(3)_S$-structures that involve the natural SO(3)-structure defined by the Einstein metric on $\Sigma_3$, with modifications induced by the $S^4$ fibration. More concretely, we can expect the ordinary SO$(3)_S$-structures to be diagonal combinations of the geometric SO(3)-structure on $\Sigma_3$ and either of two distinct SO(3) subgroups (\ref{eq:SO5subgroups}) of the SO(5) isometry that acts on the fibred $S^4$. In ordinary geometry, these structures must descend from the SO(7)-structure defined by the conventional metric. Altogether, we have the following branchings:
\begin{equation} \label{eq:GroupTh}
\textrm{SO}(7) \supset
\textrm{SO}(4) \times \textrm{SO}(3) \sim
\textrm{SO}(3)_- \times\textrm{SO}(3)_+ \times \textrm{SO}(3) \supset
\hspace{-8pt}
\begin{array}{c}
\textrm{SO}(3)^\prime \times \textrm{SO}(3) \supset \textrm{SO}(3)_S \; $\textrm(SLAG)$, \\[6pt]
\textrm{SO}(3)_+ \times  \textrm{SO}(3)_S^\prime \supset \textrm{SO}(3)_S^\prime  \; $\textrm(A3C)$.
 \end{array}
\end{equation}
The SO$(3)$ that commutes with SO$(4)$ inside SO$(7)$ in the first step of (\ref{eq:GroupTh}) can be identified with the structure group of $\Sigma_3$. In the SLAG case, the $\textrm{SO}(3)^\prime$ that appears in the upper row of (\ref{eq:GroupTh}) is the diagonal of $\textrm{SO}(3)_- \times\textrm{SO}(3)_+$, and then $\textrm{SO}(3)_S$ is the further diagonal of $\textrm{SO}(3)^\prime \times \textrm{SO}(3)$. In the A3C case, $\textrm{SO}(3)_S^\prime$ is the diagonal of $\textrm{SO}(3)_- \times \textrm{SO}(3)$. 
 
These ordinary SO$(3)_S$ and SO$(3)_S^\prime$-structures give rise to generalised SO$(3)_S$ and SO$(3)_S^\prime$-structures in the ExGG description of the configurations at hand. These can be seen as reductions of the generic SU$(8)$ generalised structure via the inclusion
\begin{equation} \label{eq:SO7inSU8}
\textrm{SU}(8) \supset \textrm{SO}(8) \supset \textrm{SO}(7) \; , 
\end{equation}
followed by (\ref{eq:GroupTh}). Equivalently, SO$(3)_S$ and SO$(3)_S^\prime$ can be embedded into SU$(8)$ through $\textrm{SO}(5) \times \textrm{SO}(3)$, a route more natural in ExGG than the SO(7) route (\ref{eq:SO7inSU8}), (\ref{eq:GroupTh}). In the SLAG case, we may consider
\begin{equation} \label{eq:SO3SBreaking}
\textrm{SU}(8) \; \supset \; 
\textrm{SO}(8) \; \supset \; 
\textrm{SO}(5) \times \textrm{SO}(3) \; \supset \;  
\textrm{SO}(2) \times \textrm{SO}(3)^\prime \times \textrm{SO}(3) \; \supset \; 
\textrm{SO}(3)_S \; ,
\end{equation}
and in the A3C case,
{\setlength\arraycolsep{2pt}
\begin{eqnarray} \label{eq:SO3SA3CBreaking}
\textrm{SU}(8)  & \; \supset \; &
\textrm{SO}(8) \; \supset \; 
\textrm{SO}(5) \times \textrm{SO}(3) \; \supset \;  
\textrm{SO}(4) \times \textrm{SO}(3)  \; \sim \; 
\textrm{SO}(3)_- \times\textrm{SO}(3)_+ \times \textrm{SO}(3) \nonumber \\[4pt]
& \; \supset \; & \textrm{SO}(3)_S^\prime \times \textrm{SO}(3)_+  \; \supset \;  
\textrm{SO}(3)_S^\prime \; .
\end{eqnarray}
}The subgroups $\textrm{SO}(3)^\prime$ and $\textrm{SO}(4) \times \textrm{SO}(3)$ that respectively appear in (\ref{eq:SO3SBreaking}) and (\ref{eq:SO3SA3CBreaking}) are the same than those that appear in (\ref{eq:GroupTh}). Of course, as emphasised above, these generalised SO$(3)_S$-structures can also be seen as enlargements of the GIS defined in either configuration of interest, SLAG and A3C.

The subgroups (\ref{eq:SO3SBreaking}), (\ref{eq:SO3SA3CBreaking}) were also considered in \cite{Pico:2025cmc} as invariance groups of two different consistent subsectors of $D=4$ $\cN=8$ TCSO$(5,0,0;\mathrm{V})$-gauged supergravity. In the next subsections we will work out the explicit $D=11$ uplift of these $D=4$ subsectors by particularising our maximal truncation of section \ref{sec:CTFrom11D} to the $\textrm{SO}(3)_S$ and $\textrm{SO}(3)_S^\prime$-invariant sectors. In the language of \cite{Cassani:2019vcl,Josse:2025uro}, this corresponds to truncating $D=11$ supergravity on the $\Sigma_3 \rtimes S^4$ manifolds equipped with the generalised $\textrm{SO}(3)_S$ and $\textrm{SO}(3)_S^\prime$-structures described above. As emphasised in \cite{Pico:2025cmc}, neither SO$(3)_S$ nor SO$(3)_S^\prime$ are contained in the gauge group, TCSO$(5,0,0;\mathrm{N})$ of the full $D=4$ $\cN=8$ gauged supergravity. Nevertheless, both SO$(3)_S$ and SO$(3)_S^\prime$ are invariance groups in ExGG, and this must have geometric consequences. Indeed, as we will see, the SO$(3)_S$ and SO$(3)_S^\prime$-invariant truncations will retain only $D=4$ supergravity fields that preserve the $\Sigma_3$ submanifold of $\Sigma_3 \rtimes S^4$ without inducing deformations on it.


\subsection{Special lagrangian subsector} \label{sec:SLAG}


Following \cite{Pico:2025cmc}, we will refer to the SO$(3)_S$-invariant subsector of the $\cN=8$ TCSO$(5,0,0;\mathrm{V})$ supergravity as the SLAG subsector, as it contains an $\cN=2$ vacuum that uplifts to the SLAG solution of \cite{Gauntlett:2006ux}. This subsector corresponds to a $D=4$ $\cN=2$ supergravity coupled to one vector multiplet and two hypermultiplets, and was first obtained from straight reduction from $D=11$ in \cite{Donos:2010ax}, without passing from the intermediate $D=4$ $\cN=8$ TCSO$(5,0,0;\mathrm{N})$ supergravity. Here, we will recover the truncation of \cite{Donos:2010ax} as a subsector of our $\cN=8$ truncation, making contact with the formalism of~\cite{Cassani:2019vcl,Josse:2025uro} along the way. For simplicity, we will focus on the further subsector containing gravity and the three dilatons, $\varphi_0$, $\varphi_2$, $\varphi_3$, of the $\cN=2$ scalar manifold. This model is described by the Lagrangian \cite{Pico:2025cmc}
\begin{equation}	\label{eq:LagrangianSLAG}
{\cal L} =   R \, \textrm{vol}_4 + \tfrac32 (d\varphi_0)^2  + \tfrac12 (d\varphi_1)^2  +  \tfrac32 (d\varphi_2)^2 - V \, \textrm{vol}_4  \;  ,
\end{equation}
where $(d\varphi_0)^2 \equiv d\varphi_0 \wedge * d\varphi_0$, etc., and the scalar potential reads
{\setlength\arraycolsep{2pt}
\begin{eqnarray} \label{eq:PotSLAG}
V &= & \tfrac18 \, g^2 \, \Big(
    4 e^{3 \varphi_0 - \varphi_1 - 3 \varphi_2}
    - 12 e^{-\varphi_0 + 3 \varphi_1 - 3 \varphi_2}
    - 24 e^{\varphi_0 - \varphi_1 -  \varphi_2}
    - 8 e^{3 \varphi_0 }
    - 24 e^{ \varphi_0  + 2 \varphi_2} \Big) ,
\end{eqnarray}
}where $g \equiv g_1 = \sqrt{2} \, g_2$. For later reference, this model attains an AdS vacuum at \cite{Pico:2025cmc}
{\setlength\arraycolsep{0pt}
\begin{eqnarray} \label{eq:SLAGvac}
& \textrm{SLAG: } \quad e^{2\varphi_0} = 2 \; , \quad 
\varphi_1 = \varphi_2 = 0 \; , \qquad 
L^2=  \sqrt 2 \,  g^{-2}  \,.
\end{eqnarray}
}The AdS radius, $L^2 = -6/V$ with $V$ in (\ref{eq:PotSLAG}) evaluated on the scalar values in (\ref{eq:SLAGvac}), has also been included here. Within the full $D=4$ $\cN=8$ supergravity, this vacuum preserves $\cN=2$ supersymmetry and $\mathrm{U}(1)_R \sim \textrm{SO}(2)$ R-symmetry, with the latter identified with the SO(2) factor in (\ref{eq:SO3SBreaking}) \cite{Pico:2025cmc}. See also \cite{Pico:2025cmc} for the dictionary between the fields employed in (\ref{eq:LagrangianSLAG}), (\ref{eq:PotSLAG}) and in \cite{Donos:2010ax}. 

Let us now discuss the embedding of the $D=4$ subsector (\ref{eq:LagrangianSLAG}), (\ref{eq:PotSLAG}) into $D=11$, following the ExGG formalism of \cite{Cassani:2019vcl,Josse:2025uro} combined with our $\cN=8$ results. We start by considering the generalised SO$(3)_S$-structure on $G_3 \rtimes S^4$ described in section~\ref{sec:SO3S}, and consider the SO$(3)_S$-invariant generalised tensors that descend from the relevant generalised bundles in the $\bm{56}_{1}$ and $\bm{133}_{0}$ of $\mathbb{R}^+ \times \textrm{E}_{7(7)}$. At this point, we can take advantage of the fact that $G_3 \rtimes S^4$ actually has a GIS in order to express the former tensors as combinations of the latter. In particular, there are four SO$(3)_S$-invariant generalised vectors, ${\cal E}_I{}^M$, $I=1, \ldots ,4$, descending from the $\bm{56}_{1}$, which can be written as
\begin{equation} \label{eq:SLAGInvVec}
\hat{{\cal E}}_I{}^M (y) = \hat{K}_I{}^{\uN} \, \hat{\tilde{E}}^M{}_{\uN} (y) \; .
\end{equation}
Here, $\hat{\tilde{E}}^M{}_{\uN} (y)$ is the twisted GIS frame on $G_3 \rtimes S^4$ given in (\ref{eq:TwistedFrame}), and $\hat{K}_I{}^{\uN}$ are constant SO$(3)_S$-invariants with only non-vanishing components \cite{Pico:2025cmc}
\begin{equation} \label{eq:SLAGInvVecs}
\hat{K}_1{}^{a \overline{b}} = \tfrac{1}{\sqrt 3} \, \delta^{a \overline{b}} \; , \qquad
\hat{K}_2{}^{\alpha \beta} = \epsilon^{\alpha \beta} \; , \qquad
\hat{K}_{3 \, a \overline{b}} = \tfrac{1}{\sqrt 3} \, \delta_{a \overline{b}} \; , \qquad
\hat{K}_{4  \, \alpha \beta} = \epsilon_{\alpha \beta} \; .
\end{equation}
The generalised vectors (\ref{eq:SLAGInvVec}) close among themselves under the action of the generalised Lie derivative (\ref{eq:LieDer}) with vanishing coefficients. This leads to an abelian gauging upon truncation to $D=4$, in agreement with \cite{Donos:2010ax}, with well-defined gauge field strengths. Had we instead used the direct product frame (\ref{eq:GenFrameSplit2}), (\ref{eq:GenFrameG3S4}) in the definition (\ref{eq:SLAGInvVec}) of the SO$(3)_S$-invariant generalised vectors $\hat{{\cal E}}_I{}^M$, we would have found inconsistencies in the gauge field strengths along the lines discussed in \cite{Pico:2025cmc}.

Finally, the explicit $D=11$ embedding of this sector can be obtained by restricting the generalised Scherk-Schwarz expressions (\ref{eq:GenSS}) to the SO$(3)_S$-invariant directions (\ref{eq:SLAGInvVec}), and then writing out the ExGG fields into their GL$(7)$-covariant components (\ref{eq:ExObjects}). For the $D=11$ metric in particular we find, using (\ref{eq:Frames_UpliftExpressions}),
\begin{equation} \label{eq:SLAGMetUplift}
ds_{11}^2 = e^{2 \tilde \Delta} ds_{4}^2 + ds_{7}^2 \; , 
\end{equation}
with warp factor
\begin{equation} \label{eq:SLAGWarping}
\begin{split}
e^{2 \tilde \Delta} =e^{\frac{1}{3}(\varphi_0 - 2\varphi_1)}
\Big(e^{2\varphi_2} \cos^2\theta + e^{2\varphi_0} \big( e^{\varphi_1+3\varphi_2} \cos^2\psi + \sin^2\psi \big) \sin^2\theta
\Big)^{\frac{1}{3}} \; ,
\end{split}
\end{equation}
and internal metric
{\setlength\arraycolsep{2pt}
\begin{eqnarray} \label{eq:11DSLAGMet}
ds_{7}^2 & = &  g^{-2}   e^{-4 \tilde \Delta} \Big[
 e^{-\varphi_0+\varphi_1-\varphi_2} e^{6 \tilde \Delta}g_{\Sigma} +  \Big(
e^{3\varphi_2- \varphi_1 } \cos^2\psi + e^{-2 \varphi_1 } \sin^2\psi
\Big) \sin^2\theta \, d\psi^2 \nonumber  \\[4pt]
 && +  \Big(
e^{2\varphi_0-\varphi_1 + \varphi_2}  \sin^2 \theta + \big( e^{-2 \varphi_1 } \cos^2\psi + e^{3\varphi_2-\varphi_1} \sin^2\psi \big) \cos^2\theta
\Big) d\theta^2  \\[4pt] 
&&  + (e^{3\varphi_2-\varphi_1} - e^{-2 \varphi_1 } )  \sin 2\theta \,  \sin 2\psi  \, 
d\theta d\psi
+ e^{2\varphi_0+\varphi_1+\varphi_2} \cos^2\theta \, 
\delta_{ab} \,  D\tilde{\mu}^a   D\tilde{\mu}^b
\Big] \; . \nonumber 
\end{eqnarray}
}Here, we have set $R_3 = \sqrt{2} \, R \equiv \sqrt{2} \, g^{-1} $. The $S^4$ angles $\theta$, $\psi$ and the quantities $\tilde{\mu}^a$ are defined in (\ref{eq:SLAGangles}). The latter have covariant derivatives $D\tilde{\mu}^a = d\tilde{\mu}^a + \omega^a{}_b \, \tilde{\mu}^b$ inherited from the leftmost relation in (\ref{eq:CovDers}). Finally, $g_\Sigma = \delta_{ab} e^a e^b$ is the local metric (\ref{eq:H3MetricUHP}) on the type V group manifold, $\Sigma_3 = H^3$. Unlike the metric on $S^4$, the metric $g_\Sigma$ appears undeformed by the $D=4$ scalars, as already advertised. While we have only included some of them here for simplicity, the full $D=4$ SO$(3)_S$-invariant sector is mindful of $g_\Sigma$ and does not inflict deformations on it. For the retained fields, (\ref{eq:11DSLAGMet}) makes explicit at the level of the $D=11$ metric the consistent embedding discussed in \cite{Donos:2010ax}.

On the $\cN=2$ vacuum (\ref{eq:SLAGvac}), the metric (\ref{eq:SLAGMetUplift}) evaluates to
{\setlength\arraycolsep{2pt}
\begin{eqnarray} \label{eq:SLAGMetUpliftAdS}
ds_{11}^2 & = & L^2 \, e^{2 \tilde \Delta} \, ds^2(\text{AdS}_4) \\[4pt]
&&  + g^{-2}  \Big( \tfrac{1}{\sqrt{2}} e^{2 \tilde \Delta}  \, g_{\Sigma }  + \tfrac{1}{\sqrt{2}} e^{2 \tilde \Delta}   d \theta^2 + e^{-4 \tilde \Delta}  \sin^2 \theta \,  d \psi^2 + 2 e^{-4 \tilde \Delta} \cos^2 \theta \, \delta _{ab}  D\tilde{\mu}^a D\tilde{\mu}^b \Big) \,, \nonumber 
\end{eqnarray}
}with $ds^2(\text{AdS}_4)$ unit-radius, $L^2$ given in (\ref{eq:SLAGvac}), and warping, from (\ref{eq:SLAGWarping}),
\begin{equation}
e^{2 \tilde \Delta}=2^{\frac16} \left(\sin^2 (\theta )+1\right)^{\frac13}\,.
\end{equation}
This is the metric for the $\textrm{AdS}_4 \times (\Sigma_3 \rtimes S^4)$ near-horizon geometry for M5-branes wrapped on a SLAG three-cycle $\Sigma_3$ inside a Calabi-Yau three-fold \cite{Gauntlett:2006ux}, as given in~{\it e.g.} \cite{Gang:2014qla}.



\subsection{Associative three-cycle subsector} \label{sec:A3C}


We now turn to discuss the $D=11$ embedding of the A3C subsector of $D=4$ $\cN=8$ TCSO$(5,0,0;\mathrm{V})$ supergravity. This sector is invariant under the SO$(3)_S^\prime$ defined in (\ref{eq:SO3SA3CBreaking}), and corresponds by itself to a $D=4$ $\cN=1$ supergravity coupled to three vector multiplets, in the adjoint of the SO$(3)_+$ defined in (\ref{eq:SO3SA3CBreaking}), and two chiral multiplets \cite{Pico:2025cmc}. A further subsector of this model was first obtained from straight reduction from $D=11$ in \cite{Gauntlett:2002rv}, without passing from the intermediate $D=4$ $\cN=8$ supergravity. Here we will recover this model as a subtruncation of our $\cN=8$ truncation, combining our A3C GIS of section \ref{sec:TruncTwisted} with the generic generalised $G$-structure truncation strategy of \cite{Cassani:2019vcl}. The construction parallels that of
section \ref{sec:SLAG}, so we will now be brief.

The gravity plus dilaton, $\varphi_0$, $\varphi_1$, sector of this model is subject to the Lagrangian \cite{Pico:2025cmc}
\begin{equation} \label{eq:LagA3C}
{\cal L} =   R \, \textrm{vol}_4 + \tfrac12 (d\varphi_0)^2  + 3 (d\varphi_1)^2   - V \, \textrm{vol}_4   ,
\end{equation}
where the scalar potential is
{\setlength\arraycolsep{2pt}
\begin{equation} \label{eq:PotA3C}
V = \tfrac{1}{16} \,  g^2 
\big( 
8 e^{-\varphi_0 - 6\varphi_1}
+ 3 e^{\varphi_0 - 4\varphi_1}
+ 48 e^{\varphi_0 - 2\varphi_1}
- 64 e^{\varphi_0}
- 64 e^{-3\varphi_1} \big)
\end{equation}
}with, again, $g \equiv g_1 = \sqrt{2} \, g_2$ for simplicity. This model reaches an AdS solution at the following scalar values and with the following AdS radius $L$: 
\begin{eqnarray} \label{eq:vacA3C}
& \textrm{A3C: } \quad e^{\varphi_0} =e^{-\varphi_1} =  2 \sqrt{\frac25} \; , \quad
\chi_0 = \chi_1 = 0 \; , \quad 
L^2=\frac{25}{32} \sqrt{\frac{5}{2}}  \, g^{-2}  \, . \quad 
\end{eqnarray}
Inside the full $D=4$ $\cN=8$ TCSO$(5,0,0;\mathrm{N})$ theory, the vacuum (\ref{eq:vacA3C}) spontaneously breaks the $\cN=8$ supersymmetry down to $\cN=1$, and the gauge group down to the $\mathrm{SO}(3)_+$ subgroup defined in (\ref{eq:SO3SA3CBreaking}). Please refer to \cite{Pico:2025cmc} for the correspondence of our fields and those used in \cite{Gauntlett:2002rv}. 

The A3C subtruncation proceeds by equipping $G_3 \rtimes S^4$ with the SO$(3)_S^\prime$ generalised structure described in section \ref{sec:SO3S}, expressing the SO$(3)_S^\prime$-invariant generalised tensors in terms of the A3C GIS described in section \ref{sec:TruncTwisted}, bringing these fields to the generalised Scherk-Schwarz expressions (\ref{eq:GenSS}), and finally expressing the ExGG fields in terms of their GL(7)-covariant components (\ref{eq:ExObjects}), (\ref{eq:Frames_UpliftExpressions}) in order to make contact with the conventional $D=11$ formalism of \cite{Cremmer:1978km}. In particular, there are now $6$ invariant generalised vector fields $\hat {\cal E}_I{}^M$, $I=1, \ldots ,6$, descending from the $\bm{56}_{-1}$. These can be expressed, as in  (\ref{eq:SLAGInvVec}), in terms of the A3C twisted frames $\hat{\tilde{E}}^M{}_{\uN} (y)$ obtained in section \ref{sec:TruncTwisted}, along with suitable SO$(3)_S^\prime$-invariant constant tensors $\hat{K}_I{}^{\uN}$. We emphasise that it is crucial to use the twisted frames $\hat{\tilde{E}}^M{}_{\uN} (y)$ in this step, and not the direct product frame (\ref{eq:GenFrameSplit2}), (\ref{eq:GenFrameG3S4}) in order to avoid the inconsistencies described in \cite{Pico:2025cmc}. Altogether, the embedding of the A3C $D=4$ fields we are considering into the $D=11$ metric is 
\begin{equation} \label{eq:A3CMetUplift}
ds_{11}^2 = e^{- 2\varphi_1}
\, \bar{\Delta}^{1/3} \,  ds_{4}^2 + ds_{7}^2 \; , 
\end{equation}
where 
\begin{equation}
\bar{\Delta} \equiv
e^{\varphi_0 } \cos^2\theta +  e^{2\varphi_0 + 3\varphi_1} \sin^2\theta  \; ,
\end{equation}
and the internal metric is
\begin{eqnarray} \label{eq:A3CUplift}
ds_{7}^2
=  g^{-2} \, \bar{\Delta}^{1/3} \Big(
2
e^{-\varphi_0}
\, g_{\Sigma}
+
e^{\varphi_1}
 \, d\theta^2 
+ 
e^{\varphi_0+\varphi_1}
\bar{\Delta}^{-1}
\sin^2\theta \,
\delta_{ab} \, \mu^a \mu^b
\Big) \; .
\end{eqnarray}
%
%
%
In these expressions, we have again set $R_3 = \sqrt{2} \, R \equiv \sqrt{2} \, g^{-1} $. Also, $\theta$ is an $S^4$ angle defined in (\ref{eq:A3Cangles}), and $\mu^a \equiv \sigma^a + \frac{1}{4}\,\epsilon^{abc}\,\omega_{bc}$, with $\sigma^a$ the SU$(2)$ left-invariant forms (\ref{eq:LIS3FromS4}) defined on an $S^3$ inside $S^4$. Finally, $g_\Sigma = \delta_{ab} e^a e^b$ is again the local metric (\ref{eq:H3MetricUHP}) on the type V group manifold, $\Sigma_3 = H^3$. Again, $g_\Sigma$ is left untouched by the SO$(3)_S^\prime$-invariant $D=4$ supergravity fields. 

Evaluated on the $\cN=1$ vacuum (\ref{eq:vacA3C}), the metric (\ref{eq:A3CMetUplift}) reduces to
\begin{equation} \label{eq:A3CMetUpliftAdS}
ds_{11}^2 = 2^{\frac{5}{2}} 5^{-\frac{7}{6}} \, \Delta_0^{\frac{1}{3}} \,  L^2 ds^2(\text{AdS}_4) + \tfrac14 \, 5^{\frac{1}{3}} g^{-2} \Delta_0^{1/3} \Big( 2 g_{\Sigma} + d\theta^2 + 8 \Delta_0^{-1} \sin^2\theta \, \delta_{ab} \,  \mu^a \mu^b \Big) \; ,
\end{equation}
with
\begin{equation}
\Delta_0 \equiv
8 \cos^2\theta +  5 \sin^2\theta  \; .
\end{equation}
This is the metric for the $\textrm{AdS}_4 \times (\Sigma_3 \rtimes S^4)$ near-horizon configuration of M5-branes wrapped on an associative three-cycle $\Sigma_3$ inside a seven-dimensional manifold of G$_2$ holonomy \cite{Acharya:2000mu} as given, after a $\frac{\pi}{2}$ shift of $\theta$, in (9.79) of \cite{Gauntlett:2006ux}.


\section{Discussion} \label{sec:Discussion}


We have constructed new maximally supersymmetric consistent truncations of $D=11$ supergravity to four dimensions. The internal seven-dimensional reduction manifold, $\Sigma_3 \rtimes S^4$, arises near the AdS$_4$ horizon of the supersymmetric wrapped M5-brane configurations of \cite{Acharya:2000mu,Gauntlett:2006ux}. The M5-branes are wrapped on the SLAG or A3C cycle $\Sigma_3$, which is in turn embedded into a Calabi-Yau three-fold or a G$_2$-holonomy seven-dimensional manifold. In the construction of \cite{Acharya:2000mu,Gauntlett:2006ux}, $\Sigma_3$ needs to be diffeomorphic to $S^3$, $T^3$ or $H^3$, possibly compactified with a discrete group of isometries in the latter case. Moreover, for each of these three choices, $\Sigma_3$ must be equipped with the usual Einstein metric of positive, vanishing, or negative curvature. These three cases are diffeomorphic to the three-dimensional group manifolds $G_3$ of Bianchi types IX, I and V. Our local truncation proceeds consistently for any fibred manifold $G_3 \rtimes S^4$, with $S^4$ fibred over the group $G_3$ through the same twists as \cite{Acharya:2000mu,Gauntlett:2006ux}. For fixed $G_3$ of Bianchi type N, the resulting $D=4$ $\cN=8$ theory is the TCSO$(5,0,0;\mathrm{N})$ supergravity that we recently introduced in \cite{Pico:2025cmc}.

Only for $\Sigma_3 = H^3$ do the SLAG and A3C supersymmetric $\textrm{AdS}_4 \times (\Sigma_3 \rtimes S^4)$ configurations of \cite{Acharya:2000mu,Gauntlett:2006ux} solve the equations of motion of $D=11$ supergravity. Accordingly, $D=4$ $\cN=8$ TCSO$(5,0,0;\mathrm{V})$-gauged supergravity admits $\cN=2$ and $\cN=1$ AdS vacua \cite{Pico:2025cmc} that uplift to the eleven-dimensional solutions of \cite{Acharya:2000mu,Gauntlett:2006ux}. In any case, our $D=11$ truncations on $G_3 \rtimes S^4$ are consistent for any $G_3$. This means that any solution of $D=4$ $\cN=8$ TCSO$(5,0,0;\mathrm{N})$-gauged supergravity will give rise to a $D=11$ solution. For this reason, it would be interesting to scan for supersymmetric vacua of the TCSO$(5,0,0;\mathrm{N})$ models by employing the methods of \cite{Comsa:2019rcz,Bobev:2020qev,Berman:2022jqn}. Using our results, these vacua could be uplifted to $D=11$ supergravity, thus providing new M-theory backgrounds. In any case, our construction provides a generalisation of \cite{Acharya:2000mu,Gauntlett:2006ux} in that $\Sigma_3$ there needs to be equipped with an Einstein metric. In our case, $\Sigma_3$ can be locally taken to be any group manifold, thus of constant curvature but not necessarily Einstein. Note however, that for non-unimodular $G_3$, the reduction manifold $G_3 \rtimes S^4$ will be non-compact. In particular, the Bianchi type V truncation is only defined locally for $\Sigma_3 \rtimes S^4$ with $\Sigma_3 = H^3$. A different generalisation of \cite{Acharya:2000mu,Gauntlett:2006ux} has also been given recently in \cite{Ferrero:2020laf,Ferrero:2021wvk}, also in the context of the M5-brane. The latter construction also features an internal space of the form $\Sigma_3 \rtimes S^4$, but the fibration is no longer given by the usual topological twist of \cite{Witten:1988ze,Bershadsky:1995qy}. The wrapped manifold $\Sigma_3$ in this case no longer has constant curvature, and features orbifold singularities. It would be interesting to see if our consistent truncations can be extended to these cases.

The configurations $\textrm{AdS}_4 \times (\Sigma_3 \rtimes S^4)$ of \cite{Acharya:2000mu,Gauntlett:2006ux} were originally obtained as solutions of seven-dimensional gauged supergravity, and then uplifted on $S^4$ to $D=11$ using the consistent truncation of \cite{Nastase:1999cb,Nastase:1999kf}. It should thus be possible to obtain the $D=4$ $\cN=8$ TCSO$(5,0,0;\mathrm{N})$ supergravities from  Scherk-Schwarz reduction of $D=7$ SO(5)-gauged supergravity on the group manifold $G_3$. Alternatively, $D=11$ supergravity can be also Scherk-Schwarz reduced on $G_3$ to eight-dimensional supergravity \cite{AlonsoAlberca:2003jq,Bergshoeff:2003ri}. It would be interesting to determine if the resulting $D=8$ supergravity can be further reduced consistently on $S^4$ down to $D=4$ $\cN=8$ TCSO$(5,0,0;\mathrm{N})$-gauged supergravity. More generally, it would be interesting to determine the higher-dimensional origin of the entire family of TCSO$(p,q,r;\mathrm{N})$ supergravities of \cite{Pico:2025cmc}. This should proceed by replacing $S^4$ with a suitable hyperboloid, as in \cite{Hohm:2014qga,Baron:2014bya}. In at least two situations this is straightforward with the results that are already available. Firstly, for $p+q=5$, $r=0$ and arbitrary $G_3$, the truncation simply proceeds as the $S^4$ truncation discussed here, with $S^4$ replaced with a four-dimensional hyperboloid $H^{p+q-1}$, see appendix~\ref{sec:GISDetails}. Secondly, for general $p$, $q$, $r$ with $p+q+r=5$ and unimodular $G_3$, the gauging TCSO$(p,q,r;\mathrm{N})$ reduces to a specific type included in the dyonic CSO class of \cite{DallAgata:2011aa,Dall'Agata:2014ita}. The latter were identified in \cite{Inverso:2016eet} as descending from type IIA, or possibly $D=11$ like in our non-unimodular cases.

The duality-covariant reformulations  \cite{Coimbra:2011nw,Coimbra:2011ky,Coimbra:2012af,Berman:2010is,Hohm:2013pua,Hohm:2013uia}, like the ExGG formalism that we have used in this paper, of the higher-dimensional supergravities provide extremely powerful venues to address consistent truncations. Classification results exist in this context \cite{Inverso:2017lrz,Hassler:2022egz} for maximally supersymmetric truncations of the higher-dimensional supergravities. Yet, finding concrete examples of manifolds in those classifications that admit a constant intrinsic torsion GIS remains a challenging problem. In particular, the known maximal consistent truncations that are directly associated to holographic brane configurations of string/M-theory are still few and far between. Consistent truncations to $D=4$ $\cN=8$ supergravities were previously known related to the M2, D2 and D3-brane \cite{deWit:1986iy,Lee:2014mla,Varela:2015ywx,Guarino:2015jca,Guarino:2015vca,Ciceri:2016dmd,Cassani:2016ncu,Inverso:2016eet}. Our construction provides the first $D=4$ $\cN=8$ consistent truncation in the context of the M5-brane.


\section*{Acknowledgements}


MP is supported by predoctoral award FPU22/02084 from the Spanish Government, and partially by Spanish Government grants CEX2020-001007-S, PID2021-123017NB-I00 and PID2024-156043NB-I00, funded by MCIN/AEI/10.13039/501100011033, and ERDF, EU. OV is supported by NSF grant PHY-2310223.


\appendix

\addtocontents{toc}{\setcounter{tocdepth}{1}}



\section{Geometric conventions} \label{sec:Conventions}


Our conventions on differential geometry, and on the various manifolds, $G_3$, $S^4$, that appear in the main text are gathered in this appendix for reference.

\subsection{Differential geometry conventions} \label{sec:DGConventions}

Some standard operations in differential geometry are collected here, in order to summarise our conventions and have some relevant formulae handy for our GIS calculations. Consider an orientable Riemannian manifold $M_n$ of dimension $n$, equipped with a metric $g$. Flat, frame indices are denoted by $a,b, \dots$, and curved, coordinate indices by $m,n, \dots$. We also use the shortands $e^{a_1 \dots a_p} = e^{a_1}  \wedge e^{a_2} \wedge \dots \wedge e^{a_p}$ and $ \hat e_{a_1 \dots a_p} = \hat e_{a_1}  \wedge \hat e_{a_2} \wedge \dots \wedge \hat e_{a_p}$ for a basis of $p$-forms and polyvectors, respectively.  We define the inner product, w.r.t.~to a vector $v$, on $p$-forms by its action on a $p$-form basis as
{\setlength\arraycolsep{0pt}
\begin{eqnarray}
&& \imath_v  e^a = v^a\,,\\
&& \imath_v  e^{a_1 \dots a_p} = \imath_v e^{a_1}  \wedge e^{a_2} \wedge \dots \wedge e^{a_p}  -  e^{a_1} \wedge \imath_v e^{a_2} \wedge   \dots \wedge e^{a_p} + \dots  +(-1)^{p-1}   e^{a_1} \wedge e^{a_2}  \dots \wedge \imath_ve^{a_p} .  \nonumber
\end{eqnarray}
}On  a $p$-form $\alpha$ and a polyvector $\beta$ of order $p$, we define the $^\sharp$ and $^\flat$ operators as
\begin{equation}
(\alpha^\sharp)^{a_1 \dots a_n}= g^{a_1 b_1}\dots g^{a_n b_n} \alpha_{b_1 \dots b_n} \; ,  \qquad 
(\beta^\flat)_{a_1 \dots a_n} = g_{a_1 b_1}\dots g_{a_n b_n} \beta^{b_1 \dots b_n} \;  ,
\end{equation}
and the operators $\hat \eta$ and $\tilde \jmath_v$, with $v$ a vector, acting on a $p$-form $\alpha$ as
\begin{equation}
\hat \eta \alpha = (-1)^p \alpha \; , \qquad \tilde \jmath_v \alpha = v^\flat \wedge \alpha\,.
\end{equation}

The Hodge dual is defined by its action on the $p$-forms basis as
\begin{equation}
*_n e^{a_1 \dots a_p} = \frac{\sqrt{\textrm{det} \, g}}{(n-p)!} g^{a_1 b_1} \dots g^{a_p b_p} \epsilon_{b_1 \dots b_p c_1 \dots c_{(n-p)}} e^{c_1  \dots c_{(n-p)}} \; .
\end{equation}
Its definition can be extended to polyvectors such that it commutes with $\sharp$,
\begin{equation}
*_n \hat e_{a_1  \dots  a_p} = \frac{\sqrt{\textrm{det} \, g^{-1}}}{(n-p)!} g_{a_1 b_1} \dots g_{a_p b_p} \epsilon^{b_1 \dots b_p c_1 \dots c_{(n-p)}} \hat e_{c_1  \dots  c_{(n-p)}}\,.
\end{equation}
The following identities come in handy to deal with our GIS manipulations:
\begin{equation} \label{AppendixDiffGeo Hodge i and j identities}
*_n \imath_v = - \tilde \jmath_v *_n \hat \eta \; , \qquad 
\imath_v *_n  = *_n  \tilde \jmath_v \,  \hat \eta  \; , \qquad 
\imath_v \tilde \jmath_w +  \tilde \jmath_w \imath_v = g(v,w) \oneone \,.
\end{equation}

Finally, following \cite{Pacheco:2008ps, Coimbra:2011ky}, we define the operator 
$j^q: \Lambda^p T^*M \rightarrow \Lambda^q T^*M \otimes \Lambda^{p-q} T^*M$ on a $p$-form $\alpha$ as
\begin{equation} \label{AppendixDiffGeo jq product}
j^q e^{a_1 \dots a_p} = \tfrac{p!}{q! (p-q)!} \, e^{[a_1 \dots a_q} \otimes e^{a_{q+1} \dots a_p]}\,.
\end{equation}
Given a $p_1$-form $\alpha_1$ and a $p_2$-form $\alpha_2$ with $p_1 + p_2= (n+1)$, $j\alpha_1 \wedge \alpha_2$ coincides with the definition of \cite{Pacheco:2008ps, Coimbra:2011ky}, with the further assumption that the wedge product is always taken w.r.t.~the second factor in the tensor product \eqref{AppendixDiffGeo jq product}. This operation looks trivial when considered in isolation, as it is just a rescaling of the frame $e^{a_1 \dots a_p}$. However, it becomes a non-trivial operation when the wedge product with respect to the second factor is considered. It can be proved that 
\begin{equation}
j(\alpha_1 \wedge \alpha_2) \wedge \alpha_3 = j\alpha_1 \wedge \alpha_2  \wedge \alpha_3 + (-1)^{p_1\, p_2} j \alpha_2 \wedge \alpha_1  \wedge  \alpha_3\,,
\label{AppendixDiffGeo first j identity}
\end{equation}
where $\alpha_3$ is a $p_3$-form with $p_1 + p_2 + p_3= (n+1)$. Also, for any (identically zero) $(n+1)$-form $\alpha_1 \wedge \alpha_2$ we have
\begin{equation}
0=j(\alpha_1 \wedge \alpha_2) = j\alpha_1 \wedge \alpha_2 + (-1)^{p_1\, p_2} j \alpha_2 \wedge \alpha_1\,.
\end{equation}
%

\subsection{$T^3$, $S^3$ and $H^3$ as ordinary parallelisable spaces}
\label{sec:Appendix Global Frames for Sigma}

Our construction is valid for any three-dimensional group $G_3$ of Bianchi type N. In particular, the group manifolds of Bianchi types I (and VII$_0$), IX and V (and VII$_h$, $h>0$) are respectively diffeomorphic to the three-torus, $T^3$, the three-sphere, $S^3$, and the three-hyperboloid, $H^3$. In the latter case the global structure is different, as $H^3$ can be identified with a discrete group of isometries $\Gamma$, rendering $H^3/\Gamma$ compact, whereas the same is not possible for G$_3$ of type V (or type VII$_h$, $h>0$) because these are not unimodular. This is the reason why we have emphasised in the main text the local character of our truncations. Also, only in the above cases do the group manifolds $G_3$ admit Einstein metrics of, respectively, positive, vanishing and negative curvature.

Let $e^a$ be a left-invariant vielbein with inverse $\hat e_a$, so that
\begin{equation} \label{Appendix Global Frames for Sigma fabc}
d e^a = - \tfrac12 f_{bc}{}^a e^{bc} \; , \qquad 
[\hat e_a, \hat e_b] = f_{ab}{}^c \hat e_c\,,
\end{equation} 
with the constant anholonomic coefficients $f_{ab}{}^c$ identified with the structure constants of the associated Bianchi type. In particular,
\begin{equation} \label{eq:SCBianchiV}
\mathrm{I}: \, f_{ab}{}^c=0  \; , \qquad 
\mathrm{IX}: \, f_{ab}{}^c= \epsilon_{ab}{}^c  \; , \qquad 
\mathrm{V}: \, f_{13}{}^1 = -f_{31}{}^1 = f_{23}{}^{2}= -f_{32}{}^{2} = 1  \; ,
\end{equation}
For any $G_3$, we also have the identities
\begin{equation} \label{eq:VariousIds}
e^a \wedge e^b \wedge e^c = - R_3^{-3} \epsilon^{abc} \,  \text{vol}_3 \;, \quad 
\imath_{\hat e_c } \imath_{\hat e_b} \dd e^a = -f_{bc}{}^a  \; , \quad
d e^{ab} = R_3^{-3} \ f_{cd}{}^{[a} \epsilon^{b]cd} \, \text{vol}_3 \; ,
\end{equation}
with $R_3$ an overall scale on $G_3$, as in the main text. These identities are helpful to deal with the GIS manipulations and generalised Lie derivative calculations summarised in appendix \ref{sec:GISDetails}. The spin connection of $G_3$ with flattened one-form index is related to the structure constants via $2 \omega_{[b}{}^{a}{}_{c]} = f_{bc}{}^a$, namely,
\begin{equation}
\omega_{abc}=\tfrac12 (f_{bca} - f_{abc} + f_{acb}) \; .
\end{equation}

\subsubsection{$T^3$}

The three-torus is isomorphic to $\mathrm{U}(1) \times \mathrm{U}(1) \times \mathrm{U}(1) \cong S^1 \times S^1 \times S^1$, which is parallelisable as any Lie group. Introducing angles $\alpha$, $\beta$, $\gamma$ on each circle, the Ricci flat (and flat) metric is trivially given by
\begin{equation}
ds^2_3 = R_{3}^2 \,  ( d \alpha^2 +  d \beta^2 +  d \gamma^2)\,,
\end{equation} 
with globally defined metric vector frame defined as
\begin{equation}
\hat e_a = \left(  \partial_1, - \partial_2 , \partial_3 \right)\, .
\end{equation}
The negative sign chosen here correlates with our volume convention in (\ref{eq:VariousIds}).

\subsubsection{$S^3$} 

The three-sphere is the group manifold of SU$(2)$. A parallelisation is accordingly given by the left-invariant Maurer-Cartan forms
\begin{eqnarray} \label{eq:LIS3}
e^1 = \cos\gamma\, d\beta  + \sin\gamma\,\sin\beta\, d\alpha , \;\;
e^2 = -\sin\gamma\, d\beta + \cos\gamma\,\sin\beta\, d\alpha , \;\;
e^3 = d\gamma + \cos\beta\, d\alpha, \quad 
\end{eqnarray}
written in terms of $S^3$ angles 
$0 \le \beta \le \pi$, 
$0 \le \alpha < 2\pi$,
$0 \le \gamma < 4\pi$. The positive curvature Einstein metric is thus
\begin{equation}
ds^2_3 = R^2_{3} \, \delta_{ab} \, e^a  e^b =
R^2_{3} \, \big( d\beta^2 + \sin^2\beta d\alpha^2 +( d\gamma + \cos\beta\, d\alpha )^2 \big) \; .
\end{equation}

\subsubsection{$H^3$} 

The three-dimensional hyperboloid $H^3$ can be defined as a three-dimensional generalisation of the upper-half complex plane,
\begin{equation}
H^3= \{ (x,y,z) \in \reals^3 : z > 0 \}\,,
\end{equation} 
endowed with its natural negatively-curved Einstein metric,
\begin{equation} \label{eq:H3MetricUHP}
ds^2_3 = R_3^2 \, \frac{ dx^2  + dy^2  +dz^2 }{z^2} \; .
\end{equation}
Equivalently, $H^3 = \textrm{SO}(1,3)/\textrm{SO}(3)$ can be defined as the locus $\eta_{IJ} y^I y^J=-1$, $I=1, \ldots , 4$, in $\mathbb{R}^4$ equipped with the metric $\eta_{IJ} =\text{diag}(-1,1,1,1)$. The metric induced in $H^3$ by this embedding is
\begin{equation} \label{eq:H3MetricEmbed}
ds_{3}^2 = R_3^2 \, \eta_{IJ} d y^I  d y^J\,.
\end{equation}
The metrics (\ref{eq:H3MetricUHP}), (\ref{eq:H3MetricEmbed}) are diffeomorphic, as they can be brought into each other by the change of coordinates
\begin{equation}
x= \frac{y^2}{y^1 - y^4}\,, \quad y = \frac{y^3}{y^1 - y^4}\,, \quad z = \frac{1}{y^1 - y^4}\,.
\end{equation}
A direct and inverse vielbein for the metric (\ref{eq:H3MetricUHP}) is given by
\begin{equation} \label{AppendixfSymbolsForH3}
e^1 = \tfrac{dx}{z} \; , \quad e^2 = \tfrac{dy}{z} \; , \quad e^3 = -\tfrac{dz}{z} \; , \qquad \quad
\hat e_1 =  z \partial_x \; , \quad  \hat e_2 =  z \partial_y \; , \quad \hat e_3 =  - z \partial_z  \,.
\end{equation}
This satisfies
\begin{equation}
de^1 = -e^{13} \; , \quad 
de^2 = -e^{23} \; , \quad 
de^3 = 0 \; , \quad 
\end{equation}
in agreement with (\ref{Appendix Global Frames for Sigma fabc}) with the Bianchi V structure constants $f_{ab}{}^c$ given in (\ref{eq:SCBianchiV}). This shows the local equivalence between $H^3$ and the group manifold $G_3$ of Bianchi type V.

\subsection{$S^4$ data} \label{sec:S4Data}

Our conventions for the four-sphere are as in appendix A of \cite{Lee:2014mla}. We reproduce some of their formulae here for convenience, and derive others that will be helpful for the GIS manipulations of appendix \ref{sec:GISDetails}. 

In more generality, the round, SO$(d+1)$-invariant, Einstein metric on the  $d$-dimensional sphere $S^d$ is
\begin{equation} \label{eq:SdLineEl}
ds^2_{d}= R^2 \delta_{ij} d y^i d y^j\, , 
\end{equation}
where $i,j, \ldots = 1, \ldots, d$ are $\mathrm{SO}(d+1)$ indices raised and lowered with $\delta_{ij}$, and $y^i$ are constrained $\mathbb{R}^{d+1}$ coordinates that define $S^d$ as the locus
\begin{equation} \label{eq:SdinRd+1}
\delta_{ij} y^i y^j =1 \; .
\end{equation}
The $\mathrm{SO}(d+1)$ conformal Killing vectors and their associated one-forms are given by
\begin{equation} \label{eq:SdCKVs}
R^{-1} k_i = R (d y_i)^\sharp  \; , \qquad 
R^{-1} (k_i)^\flat = R \, d y_i  \,.
\end{equation}
The following identity holds,
\begin{equation}
\imath_{k_i}\dd y^j = R^{-2} g_d(k_i, k_j) = \delta_{ij} - y_i y_j\,, 
\end{equation}
where $g_d$ is the metric on $S^d$ with line element (\ref{eq:SdLineEl}). This can be also written
\begin{equation}
g_d= \tfrac12 v^{ij\, \flat} v_{ij}{}^{\flat} \; ,
\end{equation}
in terms of the $S^d$ Killing vectors 
\begin{equation}
v_{ij} = 2 R^{-1} y_{[i} k_{j]}\, ,
\end{equation}
with $k_i$ defined in (\ref{eq:SdCKVs}). The volume form on $S^d$ can be written as
\begin{equation}
\text{vol}_{d} = \frac{R^d}{d!} \epsilon_{i_1 \dots i_{d+1}} y^{i_1} d y^{i_2} \wedge \dots \wedge d y^{i_{d+1}}=*_d 1\,.
\end{equation}
Using \eqref{AppendixDiffGeo Hodge i and j identities} it can be shown that the volume form satisfies the identities
\begin{equation} \label{eq:volS4Id}
R*_d  d y_i =  R^{-1} \imath_{k_i} \text{vol}_{d}=-\frac{R^{d-1}}{(d-1)!} \epsilon_{i l_1l_2 \dots l_d} \, y^{l_1}  d  y^{l_2}\wedge \dots \wedge d y^{l_d}\; ,
\end{equation}
and
\begin{equation}
R^2 *_d \, (d y_i \wedge d y_j) = - R^{-2} \imath_{k_i} \imath_{k_j} \text{vol}_{d} = \frac{R^{d-2}}{(d-2)!} \epsilon_{i j l_1l_2 \dots l_{d-1}} \, y^{l_1}  d  y^{l_2}\wedge \dots \wedge d y^{l_{d-1}} \; .
\end{equation}
Using \eqref{AppendixDiffGeo first j identity} in (\ref{eq:volS4Id}) one can further derive:
\begin{equation} \label{eq:Sidentity}
jR*_d d y_i  \wedge \alpha = -\frac{R^{d-2}}{(d-2)!} \tfrac12 \epsilon_{ill'l_1 \dots l_{d-2}} v^{ll'\, \flat} \otimes d y^{l_1}\wedge \dots \wedge d y^{l_{d-2}} \wedge \alpha \,,
\end{equation}
for any $[\text{dim} (\Sigma \times S^d)-2]$-form $\alpha$ on the product of $S^d$ and a manifold $\Sigma$. 

Further identities are helpful for the generalised Lie derivative manipulations of our GIS summarised in appendix \ref{sec:GISDetails}. Using the last equation in \eqref{AppendixDiffGeo Hodge i and j identities} to permute $\imath$ and $\tilde \jmath$ until the latter hits $\text{vol}_{d}$, it can be shown that
\begin{equation}
R^3 \dd y_i \wedge *_d \, (d y_j \wedge d y_k)= - R^{-3} \tilde \jmath_{k_i} \imath_{k_i} \imath_{k_j} \text{vol}_{S^d}  =2 \delta^l_{[j} \delta_{k]i} R *_d dy_l + y_i \, \imath_{v_{jk}} \text{vol}_{d}\,.
\end{equation}
One can also derive:
\begin{eqnarray}
& \imath_{v_{jk}} R *_d d y_i  - y_i R^2 *_d( d y_j \wedge d y_k ) = R^{-2}\big( y_i \imath_{k_j} \imath_{k_k} \text{vol}_{d} + y_j \imath_{k_k} \imath_{k_i} \text{vol}_{d} + y_k \imath_{k_i} \imath_{k_j}  \text{vol}_{d}  \big) \nonumber \\[4pt]
 &= -(-1)^{d} \frac{R^{d-2}}{(d-2)!} \epsilon_{l_1 \dots l_{d-2}ijk} d y^{l_1} \wedge \dots d y^{l_{d-2}}\,.
\end{eqnarray}
To prove this identity one needs to use $y_{[i} \, \epsilon_{jk l_1 \dots l_{d-1}]} = 0$ and $y_i d y^i =0$. 

Finally, fixing $d=4$, the local three-form $A$ introduced in the main text such that $dA = 3 R^{-1} \, \vol_4$ is further required to obey the algebraic constraint
\begin{equation} \label{eq:AalgCond}
jA \wedge i_{v_{jk}}R*_4 \dd y_i \wedge \text{vol}_\Sigma + jR*_4 \dd y_i \wedge i_{v_{jk}} A \wedge \text{vol}_\Sigma +jA \wedge y_i i_{v_{jk}}A \wedge \text{vol}_\Sigma =0\, ,
\end{equation}
for our GIS to have constant intrinsic torsion as derived in appendix~\ref{sec:GISDetails}. We have verified that (\ref{eq:AalgCond}) is always identically satisfied for various gauge choices of $A$.

Introducing explicit angles on $S^4$ helps simplify some calculations, particularly those involving the explicit uplifts of section~\ref{sec:Subsectors} of the main text. For the SLAG subsector uplift of section \ref{sec:SLAG}, the following angles, taken from (B.1) of \cite{Cassani:2019vcl}, are useful
\begin{equation} \label{eq:SLAGangles}
\begin{array}{lll}
y^1 & =& \sin \theta \cos \psi \, \\
y^2 & =& \sin \theta \sin \psi \,, \\
y^3 & =& \cos \theta \tilde{\mu}^3 = \cos \theta \cos \chi \,, \\
y^4 & =& \cos \theta \tilde{\mu}^4 = \cos \theta \sin \chi \cos \phi \,, \\
y^5 & =& \cos \theta \tilde{\mu}^5 = \cos \theta \sin \chi \sin \phi\, . 
\end{array}
\end{equation}
The $S^4$ angles $(\theta , \psi , \chi , \phi )$ that appear here are the same that feature in (\ref{eq:11DSLAGMet}), (\ref{eq:SLAGMetUpliftAdS}). For the A3C uplift, it is more helpful to chose instead the following parametrisation, which agrees with (D.11) of \cite{Cassani:2020cod} after some redefinitions,
\begin{equation} \label{eq:A3Cangles}
\begin{array}{lll}
y^1 &= \cos\theta, \\
y^2 &= \sin\theta\,\cos\frac{\psi}{2}\,\cos\frac{\chi+\phi}{2}, \\
y^3 &= \sin\theta\,\cos\frac{\psi}{2}\,\sin\frac{\chi+\phi}{2}, \\
y^4 &= \sin\theta\,\sin\frac{\psi}{2}\,\cos\frac{\chi-\phi}{2}, \\
y^5 &= \sin\theta\,\sin\frac{\psi}{2}\,\sin\frac{\chi-\phi}{2}.
\end{array}
\end{equation}
These are the angles used in (\ref{eq:A3CUplift}), (\ref{eq:A3CMetUpliftAdS}). In particular the SU$(2)$ left-invariant one-forms $\sigma^a$ used in those equations are, similarly to (\ref{eq:LIS3}),
{\setlength\arraycolsep{2pt}
\begin{eqnarray} \label{eq:LIS3FromS4}
\sigma^1 &= & \tfrac12 \big( \cos \psi \, d\chi + d\phi \big) \;  , \nonumber  \\[3pt]
\sigma^2 &= & \tfrac12 \big( \cos \phi \, d\psi + \sin \phi \sin \psi \, d \chi  \big) \; , \\[3pt]
\sigma^3 &= & \tfrac12 \big( - \sin \phi \, d\psi + \cos \phi \sin \psi \, d \chi  \big) \; .\nonumber 
\end{eqnarray}
}



\section{Details on the generalised identity structure} \label{sec:GISDetails}


Let us now collect some details of the derivation of the GIS's discussed in section \ref{sec:CTFrom11D} of the main text, building on the geometric data summarised in appendix \ref{sec:Conventions}.


\subsection{$\mathrm{E}_{7(7)}$ conventions} \label{sec:E7Conv}


We follow the $\mathrm{E}_{7(7)}$ conventions of \cite{Josse:2025uro}, which agree with those of \cite{Guarino:2015qaa} up to some rescalings of the generators. In particular, we use the SL$(8,\mathbb{R})$ basis in which the $\mathrm{E}_{7(7)}$ generators split as $t_\alpha = (t_A{}^B ,  t_{ABCD} )$ via $\bm{133} \rightarrow \bm{63} + \bm{70}$, where $A=1 , \ldots , 8$ labels the fundamental representation of SL$(8,\mathbb{R})$, as in the main text. We take these generators in the fundamental representation of $\mathrm{E}_{7(7)}$ as
\begin{eqnarray} \label{t-133}
&& (t_A{}^B)^{\uM}{}_{\uN}  = \begin{pmatrix}
 (t_A{}^B)^{CD}{}_{EF} &0 \\
0 & (t_A{}^B)_{CD}{}^{EF}
\end{pmatrix} =
\begin{pmatrix}
4 \delta^{[C}{}_{[E}(t_A{}^B)^{D]}{}_{F]} &0 \\
0 & -4 \delta^{[E}{}_{[C}(t_A{}^B)^{F]}{}_{D]}
\end{pmatrix} \, , \; \nonumber \\[10pt]
&& (t_{ABCD})^{\uM}{}_{\uN} = \begin{pmatrix}
  0 & (t_{ABCD})^{EFGH} \\
(t_{ABCD})_{EFGH} & 0
\end{pmatrix} = 
\begin{pmatrix}
0 & 4! \delta^{EFGH}_{ABCD} \\
\epsilon_{ABCDEFGH} & 0
\end{pmatrix} , \; 
\end{eqnarray}
and define
\begin{equation}
(t_\alpha)_{\uM}{}^{\uN}=-(t_\alpha)^{\uN}{}_{\uM} \; .
\end{equation}
In (\ref{t-133}), 
$
(t_A{}^B)^C{}_D= \delta_A^C \delta_D^B - \tfrac18 \delta_A^B \delta_D^C
$
 are the SL(8) generators in the fundamental.


\subsection{Fundamental and adjoint bundles of $\mathrm{E}_{7(7)}$ ExGG for M-theory}


Denoting by $E_0$ the vector space spanned by the $\rep{56}$ of $\Ex{7}$, by $F$ the fundamental representation of $\GLg{7}$, and by $\Lambda^p F^*$ the totally antisymmetric $p$-index representation of $\GLg{7}$, we have the branching
\begin{equation} \label{eq:E7FunSplit}
E_0 = F \oplus \Lambda^2F^*\oplus \Lambda^5F^* \oplus \left( F^* \otimes \Lambda^7F^* \right)\, ,
\end{equation}
replicating (\ref{eq:56Branching}). A generalised vector $V^M$ accordingly splits in terms of $\mathrm{E}_{7(7)} \supset \textrm{SL}(8) \supset \textrm{GL}(7)$-covariant components as
\begin{equation}
V^M= (  V^{AB} , V_{AB} ) = 
( V^{m 8} , 
 V^{m n} ,
 V_{m8} ,
 V_{m n} ) \; ,
\end{equation}
with $A=1 , \ldots ,8 $ as above, $m=1 , \ldots , 7$ GL$(7)$ indices as in the main text, and components
\begin{eqnarray}
V^{m8}=v^m \,, \quad 
V^{mn}=\tfrac{1}{5!} \epsilon^{mn p_1 \dots p_5} \sigma_{p_1 \dots p_5} \,, \quad
V_{mn}=\omega_{mn} \,, \quad
V_{m8}=\tfrac{1}{7!} \epsilon^{n_1 \dots n_7} \tau_{m,n_1 \dots n_7}  \, , 
\end{eqnarray}
written in terms of objects of ordinary geometry: a vector $v$, a two-form $\omega$, a five-form $\sigma$ and a mixed symmetry tensor $\tau$, like in (\ref{eq:GenFrameSplit}) of the main text.

Under $\GLg{7}$, the $\mathbb{R}^+ \times \Ex{7} $ adjoint representation branches as
\begin{equation}
\rep{1}_0 + \rep{133}_0 = \mathbb{R}\oplus \left( F \otimes F^* \right) \oplus \Lambda^3F^*  \oplus \Lambda^3F  \oplus \Lambda^6F^*  \oplus \Lambda^6F\,.
\label{PreliminariesE77_Mth_E7xRplusGenerators}
\end{equation}
Explicitly, the $\mathbb{R}^+ \times \Ex{7} $ action of \cite{Coimbra:2011ky} on a generalised vector $V^M$ is reproduced by the Lie algebra element 
\begin{equation} \label{eq:adj}
\begin{split}
&R=l t_0 + \Big( r^m{}_n t_m{}^n + r^m{}_m \left( t_8{}^8 - \tfrac12 t_0 \right) \Big) + \tfrac{1}{4!} \tfrac{1}{3!} a_{m_1 m_2 m_3} \epsilon^{m_1 \dots m_7} t_{m_4 \dots m_7}  \\
&\quad \quad \quad + \tfrac{1}{3!} \alpha^{m_1 m_2 m_3} t_{m_1 m_2 m_3 8} - \tfrac{1}{6!} \tilde a_{m_1 \dots m_6} \epsilon^{m_1 \dots m_7} t_{m_7}{}^8 +  \tfrac{1}{6!} \tilde \alpha^{m_1 \dots m_6} \epsilon_{m_1 \dots m_7} t_8{}^{m_7}\,,  
\end{split}
\end{equation}
which effectively defines the branching \eqref{PreliminariesE77_Mth_E7xRplusGenerators}. For simplicity, we have suppressed fundamental $\Ex{7}$ indices in (\ref{eq:adj}). The algebra of $\Ex{7}$ is recovered by setting $l=\tfrac12 r^m{}_m$. It is straightforward, albeit tedious, to check that the commutator of two adjoint elements, $R''=[R,R']$ defined by (\ref{eq:adj}) verifies the same relations as in \cite{Coimbra:2011ky}. It is also useful to note that the adjoint projection $E_0 \times E_0^* \rightarrow \text{ad}_{\mathbb{R}^+ \times \Ex{7}}$ of \cite{Coimbra:2011ky} is given in $\Ex{7}$ covariant indices by the projector (\ref{eq:adtimes}) of the main text. With that definition, the generalised Lie derivative given in \cite{Hohm:2013uia} for two vectors $V$, $W$ of weights $\lambda(V)=\lambda(W)=\tfrac{1}{2}$ in exceptional field theory,
\begin{equation}
\begin{split}
L_V W^M&= V^N \partial_N W^M - \alpha \mathbb{P}^M{}_N{}^P{}_Q \partial_P(V^Q) W^N + \lambda (W) \partial_N (V^N) W^M \\
&= V^N \partial_N W^M - (\partial \times_{\text{ad}}V)_N{}^M W^N\,,
\end{split}
\end{equation}
coincides with the ExGG generalised Lie derivative of \cite{Coimbra:2011ky} when the $M$-theory section condition $\partial_M = (\partial_m,0,0,0)$ is imposed.

\subsection{Generalised parallelisations for the direct product $\Sigma_3 \times S^4 $}
\label{Appendix Generalized Parallelisations}

The E$_{7(7)}$ generalised tangent bundle for M-theory is isomorphic to \cite{Coimbra:2011ky}
\begin{equation} \label{eq:E7TanBun}
E \cong TM \oplus \Lambda^2 T^*M  \oplus \Lambda^5 T^*M \oplus (T^*M \otimes \Lambda^7 T^* M)\, ,
\end{equation}
paralleling (\ref{eq:E7FunSplit}). Our seven-dimensional internal manifold $M$ further splits as 
$
M = \Sigma_3 \times S^4 
$
with $\Sigma_3$ given, at this point, by an arbitrary Riemannian three-dimensional manifold. For such factorised manifold $M = \Sigma_3 \times S^4 $, (\ref{eq:E7TanBun}) further branches as
\begin{equation}
\begin{split}
E\,\cong \, &T\Sigma_3  \oplus \left( N'_4 \otimes  \Lambda^2 T^* \Sigma_3 \right) \oplus \left( T^* \Sigma_3 \otimes (\Lambda^4 T^* S^4 \otimes \Lambda^3 T^* \Sigma_3)  \right)\\
&\quad  \oplus  \left( \left[ \Lambda^2 T^* S^4 \oplus (T^*S^4 \otimes \Lambda^4 T^* S^4) \right] \otimes \Lambda^3 T^* \Sigma_3 \right) \oplus E_4 \oplus \left( N_4 \otimes T^*\Sigma_3 \right)\,,
\end{split}
\end{equation}
where, following \cite{Lee:2014mla,Cassani:2019vcl}, we have introduced the following generalised bundles on $S^4$
\begin{equation}
E_4 \cong TS^4 \oplus \Lambda^2 T^*S^4\,,\qquad
N_4 \cong T^*S^4 \oplus \Lambda^4 T^*S^4\,,\qquad
N'_4 \cong \mathbb{R} \oplus \Lambda^3 T^*S^4\, .
\end{equation}
These admit generalised frames given by \cite{Lee:2014mla,Cassani:2019vcl}
{\setlength\arraycolsep{2pt}
\begin{eqnarray}
\hat E_{ij}^0 \;  &=& \big( v_{ij} \; , \;  R^2 *_4 ( d y_i \wedge d y_j) + \imath_{v_{ij}} A  \big) \quad \in \Gamma (E_4) \; , \nonumber \\[4pt]
\hat E'^{i} \;  &=& \big(  R \,  d y^i  \; , \;  R \, d y^i\wedge A - y^i \, \text{vol}_{4}   \big) \quad \in \Gamma (N_4) \; , \nonumber \\[4pt]
\hat E_i \;  & =&   \big(  y_i \; , \;    R *_4 \, d y_i +y_i \, A   \big)  \quad  \in \Gamma (N'_4) \; ,
\end{eqnarray}
with all quantities defined in appendix~\ref{sec:Conventions}.

We have used these building blocks, along with the direct, $e^a$, and inverse, $\hat{e}_a$, frames on $\Sigma_3$, to construct our GIS on the direct product manifold $\Sigma_3 \times S^4$. This was expressed in (\ref{eq:GenFrameSplit2}), (\ref{eq:GenFrameG3S4}) of the main text in terms of the (generalised) inverse frame vectors $\hat{E}^M{}_{\uN}$. For completeness, the dual generalised frame covectors $E_M{}^{\uN}$, normalised such that $\hat{E}^M{}_{\uN}  E_M{}^{\uP} = \delta_{\uN}^{\uP}$, are
{\setlength\arraycolsep{1pt}
\begin{eqnarray} \label{eq:DirGenFrameG3S4}
E_a &=& \big( 0 \; , \; 0 \; , \; 0 \; , \; R_3^{-1} \, \hat e_a \otimes (\text{vol}_{4}^\sharp \wedge \text{vol}_{3}^\sharp)\big) \; , \nonumber \\[5pt]
E_{ab}{}^i \;  & =&   \big( 0 \; , \;R_3^{-2} \, [y^i- \tfrac{1}{4! \sqrt{\det g_{4}}} \,  \epsilon^{mnpq} \,  (R \, dy^i \wedge A)_{mnpq} ] \, \hat e_{ab} \; , \; R_3^{-2} \, \hat e_{ab} \wedge R \left(  *_4 \dd y^i\right)^\sharp \; , \; 0 \big) \; , \nonumber \\[5pt]
E^{ij} \;  &=& \big( v^{ij\, \flat} -\tfrac{1}{4!\sqrt{\det g_{4}}} \left( j A \wedge R^2 (\dd y^i \wedge \dd y^j) \right)_{m; n_1 \dots n_4} \epsilon^{ n_1 \dots n_4} dy^m  \; , \;  R^2 \left[*_4 (\dd y_i \wedge \dd y_j)\right]^\sharp , 0 , 0 \big) , \nonumber \\[4pt]
E'^a  \; & =& \big( R_3 \, e^a \; , \;  0 \; , \;  0 \; , \;  0 \big) \; ,  \\[5pt]
E'_{ia} \;  &=& \big( 0 \; , \; R_3^{-1} \, \hat e^a \wedge \left(R^{-1} k_i - y_i (*_4 A)^\sharp \right)  \; , \; - R_3^{-1} \, \hat e^a \wedge y^i \left(   \text{vol}_{4} \right)^\sharp \; , \; 0  \big) \; , \nonumber \\[5pt]
E'_{ij}&=&  \big( 0 \; , \; 0 \; , \; (R^2  (\dd y_i \wedge \dd y_j) + *_4 \imath_{v_{ij}} A)^\sharp \wedge \text{vol}_{3}^\sharp  \; , \;  -  v_{ij} \otimes (\text{vol}_{4}^\sharp \wedge \text{vol}_{3}^\sharp) \big) \; , \nonumber
\end{eqnarray}
}with all quantities again defined in appendix~\ref{sec:Conventions}.


\subsection{Intrinsic torsion}


We now further assume that $\Sigma_3$ is diffeomorphic to a local group manifold $G_3$ of arbitrary Bianchi type N, so that (\ref{Appendix Global Frames for Sigma fabc}) holds with constant structure constants $f_{ab}{}^c$. Equipped with the definitions and conventions collected in appendix \ref{sec:Conventions}, we compute the generalised Lie derivative (\ref{eq:LieDer}) of the generalised vectors $\hat{E}^M{}_{\uN}$ in (\ref{eq:GenFrameSplit2}), (\ref{eq:GenFrameG3S4}) that define our GIS on $\Sigma_3 \times S^4$, with respect to themselves. An arduous calculation involving heavy use of the identities derived in appendix~\ref{sec:Conventions}, allows us to bring the non-vanishing contributions to the following form:
\begin{equation}
\begin{array}{lll}
L_{\hat E'_a} \hat E'_b &=& R_3^{-1} f_{ab}{}^c \hat E'_c \,,\\[6pt]
L_{\hat E'_a} \hat   E'^{ib} &=& - R_3^{-1} f_{ac}{}^b  \hat   E'^{ic} \,,\\[6pt]
L_{\hat E'_a} \hat E^{bc}{}_{i} &=& - 2 R_3^{-1} f_{a[d}{}^{[b} \delta^{c]}_{e]} E^{de}{}_{i} \,,\\[6pt]
L_{\hat E'_a} \hat E^b &=&  - R_3^{-1} (f_{ac}{}^b +f_{ad}{}^d \delta^b_c) \hat E^c \,,\\[6pt]
L_{\hat E'_a} \hat E'^{ij} &=&  - R_3^{-1} f_{ab}{}^b \hat  E'^{ij} \,,\\[6pt]
L_{\hat E'^{ia}} \hat E'_b &=&  R_3^{-1} f_{bc}{}^a \hat   E'^{ic} \,,\\[6pt]
L_{\hat E'^{ia}} \hat E_{jk} &=&  R_3^{-1} f_{bc}{}^a \delta^{li}_{jk} \hat E^{bc}{}_l  \,,\\[6pt]
L_{\hat E'^{ia}} \hat E'^{jb} &=& \tfrac12 R_3^{-1} \epsilon^{bcd} f_{cd}{}^a  \hat E'^{ij}  \,,\\[6pt]
L_{\hat E'^{ia}} \hat E^{bc}{}_{j} &=& - R_3^{-1} \delta^i_{j} f_{de}{}^a \epsilon^{bce}   \hat E^{d}  \,,\\[6pt]
L_{\hat E^{ab}{}_i} \hat E'_{c} &=& R_3^{-1} \left( f_{de}{}^{[a} \delta^{b]}_c + 2f_{cd}{}^{[a} \delta^{b]}_e \right)  \hat E^{de}{}_i - 2 R^{-1} \delta^{ab}_{cd} \delta_{ij} \hat E'^{jd}\,,\\[6pt]
L_{\hat E^{ab}{}_i} \hat E_{jk} &=& -2 R^{-1} \delta^l_{[j} \delta_{k]i} \delta^{ab}_{cd} \hat E^{cd}{}_l -   \tfrac12 R_3^{-1} f_{cd}{}^{[a} \epsilon^{b]cd} \epsilon_{ijkl_1 l_2} \hat E'^{l_1 l_2}\,,\\[6pt]
L_{\hat E^{ab}{}_i} \hat E'^{jc} &=&- R^{-1} \epsilon^{abc} \delta_{ik} \hat E'^{kj} -  R_3^{-1} \delta_{i}^j\left( \epsilon^{cef} f_{ef}{}^{[b} \delta^{a]}_d - 2 f_{de}{}^{[a} \epsilon^{b]ce} \right) \hat E^d\,,\\[6pt]
L_{\hat E^{ab}{}_i} \hat  E^{cd}{}_j &=& -2 R^{-1} \delta_{ij}  \epsilon^{cd[a} \delta^{b]}_e \hat E^e\,,\\[6pt]
L_{\hat E_{ij}} \hat  E_{kl} &=& - R^{-1} \left( \delta_{ik} \hat  E_{jl} - \delta_{il} \hat  E_{jk} + \delta_{jl} \hat  E_{ik}  - \delta_{jk} \hat  E_{il} \right) \,,\\[6pt]
L_{\hat E_{ij}} \hat  E^{ka} &=& - 2 R^{-1} \delta^k_{[i} \delta_{j]l} \hat  E^{la} \,,\\[6pt]
L_{\hat E_{ij}} \hat  E'^{ab}{}_k &=&  2 R^{-1} \delta^l_{[i} \delta_{j]k} \hat  E'^{ab}{}_l \,,\\[6pt]
L_{\hat E_{ij}} \hat  E'_{kl} &=& - R^{-1} \left( \delta_{ik} \hat  E'_{jl} - \delta_{il} \hat  E'_{jk} + \delta_{jl} \hat  E'_{ik}  - \delta_{jk} \hat  E'_{il} \right) \,.
\end{array}
\label{Appendix Generalised Parallelizations_StructureConstantAnaliticComputation}
\end{equation}
The relations (\ref{Appendix Generalised Parallelizations_StructureConstantAnaliticComputation}) show that the GIS (\ref{eq:GenFrameSplit2}), (\ref{eq:GenFrameG3S4}) does indeed have constant intrinsic torsion. Further analysis allows one to cast (\ref{Appendix Generalised Parallelizations_StructureConstantAnaliticComputation}) in the succinct form (\ref{eq:CTCon}), with constant intrinsic $X_{\uM\uN}{}^{\uP}$ given, precisely, by the DP embedding tensor (\ref{eq:ETtromboneSplit}), (\ref{eq:ETtrombone}) of $D=4$ $\cN=8$ TCSO$(5,0,0;\mathrm{N})$-gauged supergravity, expressed in the E$_{7(7)}$ basis (\ref{t-133}), and with coupling constants identified as in (\ref{eq:CouplingsRadii}). This shows the consistency of the maximally supersymmetric truncation of $D=11$ supergravity on $G_3 \times S^4$ claimed in the main text.

The frames $\hat{E}^M{}_{\uN}$ in (\ref{eq:GenFrameSplit2}), (\ref{eq:GenFrameG3S4}) can also be defined at face value on $\Sigma_3 \times H^{p+q-1}$, with $\Sigma_3=G_3$ still a local group manifold, and $H^{p+q-1}$ a four-dimensional hyperboloid. This change only entails replacing the constraint (\ref{eq:SdinRd+1}) for the $\mathbb{R}^5$ coordiantes $y^i$ with
\begin{equation} \label{eq:HdinRd+1}
\eta_{ij} y^i y^j =1 \; ,
\end{equation}
where $\eta_{ij}$ is the indefinite metric (\ref{eq:Charges2bis}), with $p+q=5$, $r=0$. The identities of appendix~\ref{sec:S4Data}, which are crucial to show consistency, still hold under (\ref{eq:HdinRd+1}) with suitable sign changes. Similar manipulations allow us to show that the frames (\ref{eq:GenFrameSplit2}), (\ref{eq:GenFrameG3S4}) still close under the generalised Lie derivative (\ref{eq:LieDer}) into expressions similar to (\ref{Appendix Generalised Parallelizations_StructureConstantAnaliticComputation}), again with suitable sign changes induced by $\eta_{ij}$. As expected, the resulting expressions can again be brought to the form (\ref{eq:CTCon}) with constant torsion $X_{\uM\uN}{}^{\uP}$, now given by the $p+q=5$, $r=0$ DP embedding tensor (\ref{eq:ETtromboneSplit}), (\ref{eq:ETtrombone}) expressed in the E$_{7(7)}$ basis (\ref{t-133}), and with the identification of constants (\ref{eq:CouplingsRadii}). Thus, $D=11$ supergravity admits a consistent truncation, at least on the direct product manifold $G_3 \times H^{p+q-1}$, to $D=4$ $\cN=8$ TCSO$(p,q,0;\mathrm{N})$-gauged supergravity, with $p+q=5$.


\bibliography{references}


\end{document}